\documentclass[oneside, reqno, 11pt, a4paper]{amsart}

\AtBeginDocument{
  \DeclareSymbolFont{AMSb}{U}{msb}{m}{n}
  \DeclareSymbolFontAlphabet{\mathbb}{AMSb}
}
\DeclareFontFamily{U}{mathx}{\hyphenchar\font45}
\DeclareFontShape{U}{mathx}{m}{n}{<-> mathx10}{}
\DeclareSymbolFont{mathx}{U}{mathx}{m}{n}
\DeclareMathAccent{\widebar}{0}{mathx}{"73}
 
\usepackage[dvipsnames]{xcolor}
\usepackage{graphicx}
\usepackage{tikz}
\usepackage{caption}
\usepackage{subcaption}
\tikzstyle{arrow} = [thick,->,>=stealth]
\usetikzlibrary{shapes.misc,matrix,fit,positioning,arrows.meta,decorations.pathreplacing,calc,shapes.geometric,arrows}
\tikzset{Matrix/.style={matrix of nodes, font=\footnotesize,text height=1pt, text depth=0.5pt, text width=8.5pt, align=center, column sep=0pt, row sep=0pt, nodes in empty cells}}

\graphicspath{{figures/}}

\usepackage[a4paper, pdftex, left=2cm, top=2cm, right=2cm, bottom=2cm]{geometry}
\usepackage[final]{pdfpages}
\usepackage{changepage}

\usepackage[foot]{amsaddr}
\numberwithin{equation}{section}

\usepackage{url}
\usepackage{hyperref}
\hypersetup{
    breaklinks=true,
    bookmarksopen=true,
    pdftitle={GridapHybrid}, 
    pdfauthor={Jordi Manyer, Santiago Badia, Jai Tushar}, 
    pdfsubject={}, 
    colorlinks=true,
    linkcolor=black,
    citecolor=blue,
    filecolor=black,
    urlcolor=blue
}
\newcommand{\secref}[1]{Section~\ref{#1}}
\newcommand{\figref}[1]{Figure~\ref{#1}}

\newcommand{\lstref}[1]{Listing~\ref{#1}}

\usepackage{csquotes}
\usepackage[backend=biber, defernumbers=true, maxbibnames=5, style=numeric-comp, isbn=false, bibencoding=utf8, safeinputenc, url=false, doi=true, giveninits=true]{biblatex}
\usepackage{mathtools}
\usepackage{mathabx}

\usepackage{float}
\usepackage{framed}
\usepackage{verbatim}
\usepackage{fancyvrb}
\usepackage{booktabs}
\usepackage{colortbl}
\usepackage{multirow}
\usepackage{cancel}
\usepackage{accents}
\usepackage{mleftright}
\usepackage{thm-restate}

\usepackage[inline]{enumitem}
\usepackage{siunitx}

\numberwithin{theorem}{section}

\usepackage[breakable]{tcolorbox}
\tcbuselibrary{theorems}
\tcbuselibrary{skins}
\tcbset{
	commonstyle/.style={
		theorem style=plain,
		enhanced jigsaw,
		fonttitle=\bfseries,
		fontupper=\itshape,
		halign=justify,
		separator sign=:,
		description delimiters none,
		description font=\bfseries, 
		terminator sign={.\hspace{0.25em}},
		arc=0mm,outer arc=0mm,
		boxrule=0pt,toprule=0pt,bottomrule=0pt,leftrule=0pt,rightrule=0pt,
		titlerule=0pt,toptitle=0pt,bottomtitle=0pt,top=0pt,
		colback=white,coltitle=black,
		boxsep=0pt, bottom=0pt, left=0pt, 
	}
}
\newtcbtheorem[]{myproblem}{Problem}%
{center, commonstyle, fonttitle=\bfseries}{pb}
\newtcbtheorem[]{mydefinition}{Definition}
{center, commonstyle, fonttitle=\bfseries}{pb}
\newtcbtheorem[]{myassumption}{Assumption}
{center, commonstyle, fonttitle=\bfseries}{pb}

\usepackage{algorithm}
\usepackage[noend]{algpseudocode}

\algnewcommand\algorithmicinput{\textbf{Input:}}
\algnewcommand\Input{\item[\algorithmicinput]}
\algnewcommand\algorithmicoutput{\textbf{Output:}}
\algnewcommand\Output{\item[\algorithmicoutput]}
\algnewcommand\algorithmicassert{\textbf{Assert:}}
\algnewcommand\Assert{\item[\algorithmicassert]}

\usepackage{makecell}
\usepackage{listings}
\usepackage{textgreek}

\DeclareCaptionType{listing}[Listing][List of Listings]

\newcommand{\juliacode}[1]{
	\includegraphics[width=\textwidth]{#1.pdf}
	\vspace*{-1em}
}

\DeclareRobustCommand{\jl}[1]{\texttt{#1}}

\usepackage{amsmath, amsfonts, amssymb, amscd, bm, mathtools}

\usepackage{acronym}
\usepackage{enumitem}

\acrodef{fe}[FE]{Finite Element}
\acrodefplural{fe}[FEs]{Finite Elements}
\acrodef{dof}[DoF]{degree of freedom}
\acrodefplural{dof}[DoFs]{degrees of freedom}
\acrodef{dg}[DG]{Discontinuous Galerkin}
\acrodef{hho}[HHO]{Hybrid High-Order}
\acrodef{hdg}[HDG]{Hybridizable Discontinuous Galerkin}
\acrodef{wg}[WG]{Weak Galerkin}
\acrodef{vem}[VEM]{Virtual Element Method}
\acrodef{rhs}[RHS]{right-hand side}
\acrodef{lhs}[LHS]{left-hand side}
\acrodef{jit}[JIT]{just-in-time}
\acrodef{pde}[PDE]{partial differential equation}
\acrodefplural{pde}[PDEs]{partial differential equations}
\acrodef{kkt}[KKT]{Karush-Kuhn-Tucker}

\newcommand{\Mh}{\mathcal{M}_{h}}
\newcommand{\Th}{\mathcal{T}_{h}}
\newcommand{\Fh}{\mathcal{F}_{h}}
\newcommand{\interior}{{\rm in}}
\newcommand{\boundary}{{\rm bd}}
\newcommand{\Fhi}{\Fh^{\interior}}
\newcommand{\Fhb}{\Fh^{\boundary}}

\newcommand{\lproj}[2]{\pi_{#1}^{#2}}
\newcommand{\rec}{\mathcal{R}_T}
\newcommand{\sym}{\mathrm{sym}}
\newcommand{\symrec}{\rec^{\sym}}
\newcommand{\vecrec}{\boldsymbol{\mathcal{R}}_T}
\newcommand{\gradrec}{\mathcal{G}_T}
\newcommand{\divrec}{\mathcal{D}_T}

\newcommand{\ul}[1]{\underline{#1}}

\newcommand{\Uh}{\ul{U}_{h}}

\newcommand{\Uhb}{U_{h}^{\partial}}
\newcommand{\vUh}{\ul{\bs{U}}_h}

\newcommand{\uh}{\ul{u}_{h}}
\newcommand{\vh}{\ul{v}_{h}}

\newcommand{\uT}{\ul{u}_{T}}
\newcommand{\vT}{\ul{v}_{T}}

\newcommand{\facesum}{\sum_{F \in \partial T}}

\newcommand{\bs}[1]{\boldsymbol{#1}}

\newcommand{\Pk}[1]{\mathbb{P}^{#1}}
\newcommand{\vecPk}[1]{\left[\mathbb{P}^{#1}\right]^d}

\newcommand{\dom}{\Omega}

\newcommand{\adj}[1]{\texttt{adj}(#1)}
\newcommand{\vertices}[1]{\texttt{vert}(#1)}

\newcommand{\nextvertex}[2]{\texttt{next}(#1,#2)}

\usepackage[normalem]{ulem}
\newcounter{corr}
\definecolor{violet}{rgb}{0.580,0.,0.827}
\newcommand{\corr}[3]{\typeout{Warning : a correction remains in page \thepage}
  \stepcounter{corr}        
	      {\color{blue}\ifmmode\text{\,\sout{\ensuremath{#1}}\,}\else\sout{#1}\fi}
              {\color{red}#2}
              {\color{violet} #3}
}

\title[]{
  A natural language framework for non-conforming hybrid polytopal methods in Gridap.jl
} 

\date{\today}

\address{$^\dagger$School of mathematics\\Monash university\\Clayton\\Victoria 3800\\Australia}
\address{$^\sharp$Center for Computation and Technology and Department of Mathematics\\Louisiana State University\\Baton Rouge\\ USA.}
\author[J. Manyer]{Jordi Manyer$^{\dagger}$}
\email{jordi.manyer@monash.edu}
\author[J. Tushar]{Jai Tushar$^{\sharp}$}
\email{jai.tushar@lsu.edu}
\author[S. Badia]{Santiago Badia$^{\dagger}$}
\email{santiago.badia@monash.edu}

\begin{document}

\begin{abstract}
Hybrid finite element methods such as hybridizable discontinuous Galerkin, hybrid high-order and weak Galerkin have emerged as powerful techniques for solving partial differential equations on general polytopal meshes. Despite their diverse mathematical origins, these methods share a common computational structure involving hybrid discrete spaces, local projection operators and static condensation. This work presents a comprehensive framework for implementing such methods within the Gridap finite element library. We introduce new abstractions for polytopal mesh representation using graph-based structures, broken polynomial spaces on arbitrary mesh entities, patch-based local assembly for cell-wise linear systems, high-level local operator construction and automated static condensation. These abstractions enable concise implementations of hybrid methods while maintaining computational efficiency through Julia's just-in-time compilation and Gridap's lazy evaluation strategies. We demonstrate the framework through implementations of several non-conforming polytopal methods for the Poisson problem, linear elasticity, incompressible Stokes flow and optimal control on polytopal meshes.
\end{abstract}

\maketitle


\section{Introduction}
Hybrid \ac{fe} methods, including \ac{hdg}, \ac{hho}, non-conforming \ac{vem} and \ac{wg} methods, have emerged as a powerful class of numerical techniques for solving \acp{pde} on general polytopal meshes \cite{Cockburn_2009,DiPietro-Droniou-2020-HHO-Book,wang2013weak,Veiga-Brezzi-Marini-Russo-2023-VEM}. These methods are particularly attractive for several reasons. First, they support arbitrary order of approximation on general polygonal and polyhedral meshes, enabling flexibility in mesh generation and local refinement and coarsening strategies. Second, they achieve optimal convergence rates while maintaining local conservation properties. Third, they naturally accommodate static condensation, allowing efficient elimination of cell-based \acp{dof} to yield reduced global systems defined only on the mesh skeleton.

Despite their diverse origins and mathematical formulations, hybrid methods share a remarkable computational structure. They all rely on discrete spaces that pair volumetric unknowns with skeletal (facet-based) unknowns, employ local projection operators computed via cell-wise linear solves, assemble global systems from local contributions and exploit static condensation to reduce system size. This structural similarity suggests that a unified software framework could support the implementation of multiple hybrid methods without requiring method-specific low-level code.

However, implementing hybrid polytopal methods in existing \ac{fe} libraries presents significant challenges. Most libraries are designed around conforming methods on simplicial or hexahedral meshes and lack native support for general polytopes. Thus, although support for hybrid methods (particularly \ac{hdg}) has grown in recent years, it is often limited to conforming simplicial and hexahedral meshes. This is for instance the case for widely used libraries such as deal.II \cite{dealII93}, MFEM \cite{mfem}, FEniCS \cite{fenics}, NGSolve \cite{ngsolve} and DUNE \cite{blatt2025distributed}. There has been, however, recent effort to extend some of these libraries to support polytopal meshes, e.g. \cite{polydeal}.

Polytopal capabilities are therefore more commonly found in specialised packages such as lymph \cite{antonietti2025lymph}, PolyDiM \cite{berrone-polydim} and VEM++ \cite{dassi2025vem++}. Furthermore, the computational patterns of hybrid methods---particularly local operator assembly and static condensation---require abstractions beyond those provided by standard \ac{fe} frameworks. As a result, researchers implementing these methods often resort to writing custom, low-level code that is difficult to maintain, extend and reuse.

This work addresses these challenges by presenting a natural language framework for non-conforming hybrid polytopal methods; the user code that discretises a given \ac{pde} is concise and high-level, closely matching the mathematical notation. We realise this through a comprehensive set of extensions to the Gridap \ac{fe} library \cite{Gridap,Verdugo2022}. Gridap is a Julia-based framework that leverages \ac{jit} compilation and multiple dispatch to achieve both performance and extensibility without the traditional two-language problem. Building on Gridap's foundation, we introduce new abstractions for polytopal mesh representation, broken polynomial spaces on arbitrary mesh entities, patch-based local assembly frameworks, local operator construction and automated static condensation.

The key contributions of this work are:
\begin{itemize}
  \item Graph-based representations for general polygons and polyhedra using cyclic graphs and rotation systems, enabling efficient storage and topological queries for arbitrary polytopes.
  \item A unified framework for constructing broken polynomial spaces on arbitrary dimensional slices of polytopal meshes, supporting discrete spaces required by \ac{dg}, \ac{hdg}, \ac{hho} and related methods.
  \item A flexible patch assembly infrastructure for integrating and solving local \ac{fe} problems on arbitrary patches of mesh entities, with applications to local projection operators and static condensation.
  \item High-level abstractions for local operators that encapsulate the pattern of computing discrete projections through cell-wise linear solves, hiding implementation complexity from users.
  \item Automated static condensation procedures that eliminate cell unknowns locally to produce reduced skeleton systems, exploiting the natural block structure of hybrid formulations.
\end{itemize}

These abstractions are demonstrated through complete implementations of \ac{dg}, \ac{hdg} and \ac{hho} methods for the Poisson problem, as well as \ac{hho} discretisations for linear elasticity, incompressible Stokes flow and optimal control. The resulting implementations are remarkably concise---typically 30-50 lines of code for a complete method---while maintaining computational efficiency through Gridap's lazy evaluation and Julia's \ac{jit} compilation.

The remainder of this paper is organised as follows. \secref{sec:motivation-poisson} presents a detailed exposition of the \ac{hho} method for the Poisson problem as a motivating example, highlighting the computational building blocks shared across hybrid methods. \secref{sec:gridap} provides a brief introduction to the Gridap library and identifies the extensions required for hybrid polytopal methods. \secref{sec:implementation} describes the implementation of these extensions, including polytopal representations, broken spaces, local assembly frameworks and static condensation. \secref{sec:examples} demonstrates the framework through numerical examples of increasing complexity. Finally, we conclude with a summary of contributions and directions for future work.

\section{A motivating example: The Poisson problem with HHO} \label{sec:motivation-poisson}

To fix ideas and demonstrate the abstractions offered by the library, we consider the following Poisson problem on a polytopal Lipschitz domain  $\dom \subset \mathbb{R}^D$ with boundary $\partial \dom$: find 
\begin{align}
  \label{eq:poisson-continuous}
    u \in H^1(\dom) \ : \ -\Delta u &= f \quad \text{in } \dom, \quad u = g \quad \text{on } \partial \dom,
\end{align}
for given source data $f \in L^{2}(\dom)$ (to avoid some technicalities in the presentation) and boundary data $g \in H^{\frac{1}{2}}(\partial \dom)$. The weak formulation is: find 
\begin{equation}\label{eq:poisson-weak}
  u \in H^1_g(\dom) \ : \ a(u,v) = l(v) \quad \forall v \in H^1_0(\dom),
\end{equation}
where $H^1_0(\dom)$ and $H^1_g(\dom)$ denote the standard Sobolev spaces with functions in $H^1(\dom)$ such that their trace on $\partial\dom$ is zero and $g$, respectively, and the bilinear and linear forms are defined as
$ a(u,v) \doteq \int_{\dom} \nabla u \cdot \nabla v, \quad l(v) \doteq \int_{\dom} f v.$

We now discretise this problem using the \ac{hho} method, closely following \cite[Chapter 2]{DiPietro-Droniou-2020-HHO-Book}. This exposition will highlight the method's key features and the computational building blocks shared across the broader family of hybrid methods.

\subsection{Hybrid discrete spaces} \label{subsec:mot-hybrid-spaces}

\begin{figure}
  \centering
  \includegraphics[width=0.6\textwidth]{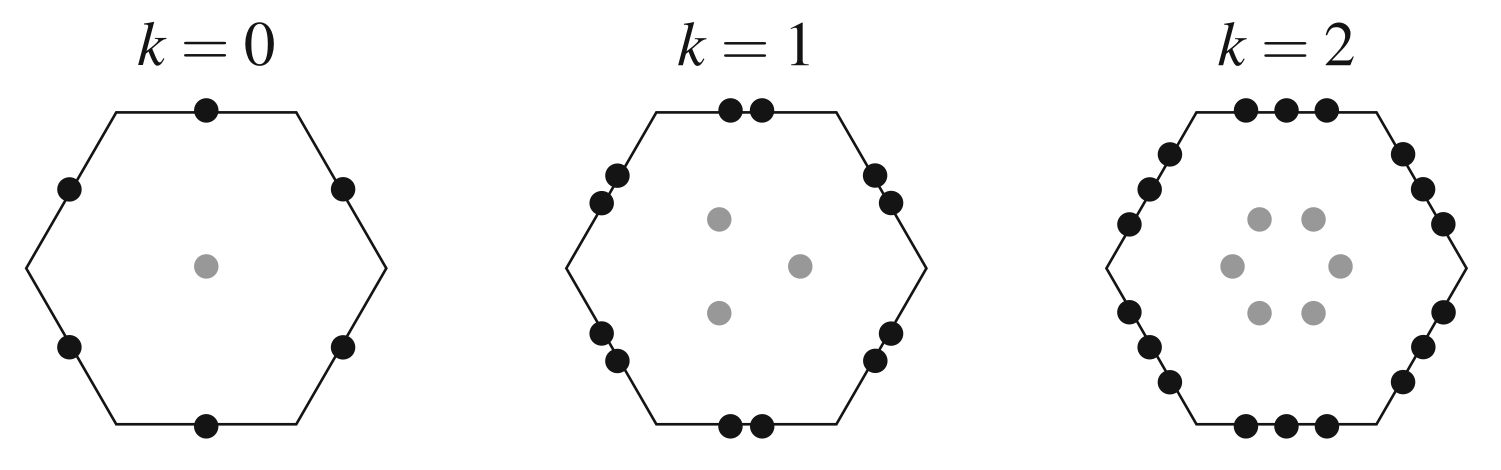}  
  \caption{Illustration of the local \ac{hho} discrete space on a hexagonal cell $T$ for polynomial degrees $(k,l)=(0,0),(1,1),(2,2)$. The space consists of polynomials defined on the cell (grey) and polynomials defined on each facet (black).}
  \label{fig:hho-space}
\end{figure}

Given a polytopal mesh $\Mh$ of $\dom$ consisting of a finite collection of non-overlapping polytopal elements $T$, we denote by $\Th$ the set of mesh cells and by $\Fh$ the set of mesh facets. For polynomial degrees $k,l \geq 0$, the \ac{hho} method introduces a \emph{hybrid} discrete space defined as

\begin{equation} \label{eq:hho-space-scalar}
  \Uh^{k,l} \doteq \Pk{k}(\Th) \times \Pk{l}(\Fh),
\end{equation}
where 
$\Pk{k}(\Th)$
and 
$\Pk{l}(\Fh)$ denote the globally discontinuous piecewise polynomial spaces of degree $k$ and $l$ on the mesh cells and facets, respectively. An element $\uh \in \Uh^{k,l}$ is thus a tuple $\uh = ((u_T)_{T \in \Th}, (u_F)_{F \in \Fh})$, where $u_T$ represents cell-wise polynomials and $u_F$ represents facet (skeleton) polynomials. This hybrid structure---pairing volumetric \acp{dof} with \acp{dof} on the mesh skeleton---is the first common feature shared by \ac{hdg}, \ac{hho}, \ac{wg} and related methods. The local spaces on each cell $T$ can be seen in \figref{fig:hho-space}.
Having defined the hybrid discrete spaces, we now describe the local operators that map between them.

\subsection{Local operators and cell-wise integration} \label{subsec:mot-local-ops}

The \ac{hho} method relies on operators defined locally on each cell $T$ and its surrounding facets. These operators are constructed by solving cell-wise linear systems that involve integrals over the cell and its facets---a second common feature of hybrid methods.

The primary building block is the \emph{local reconstruction operator} $\rec : \Uh^{k,l}(T) \to \Pk{k+1}(T)$, which computes a higher-order polynomial approximation inside $T$ from the local trace unknowns $\ul{v}_T = (v_T, (v_F)_{F \in \Fh(T)}) \in \Uh^{k,l}(T)$. This operator is uniquely defined by

\begin{align} \label{eq:hho-reconstruction}
  \int_T \nabla \rec \ul{v}_T \cdot \nabla w &= \int_T \nabla v_T \cdot \nabla w + \facesum \int_F (v_F - v_T) \nabla w \cdot \bm{n}_{TF} \quad \forall w \in \Pk{k+1}(T), \\
  \int_T \rec \ul{v}_T &= \int_T v_T,
\end{align}
where $\bm{n}_{TF}$ denotes the unit outward normal to facet $F$ from cell $T$.
Computing $\rec \ul{v}_T$ requires assembling a small local linear system involving integrals over $T$ and its facets $F \in \Fh(T)$, then solving this system to determine the coefficients of the reconstructed polynomial.

Additionally, the method employs the $L^2$-orthogonal projectors onto cells and facets, namely $\lproj{T}{0,k} : L^2(T) \to \Pk{k}(T)$ and $\lproj{F}{0,l} : L^2(F) \to \Pk{l}(F)$ such that
\begin{align} \label{eq:proj_ops}
  \int_T \lproj{T}{0,k} w \, v = \int_T w \, v \quad \forall v \in \Pk{k}(T), \quad \text{and} \quad
  \int_F \lproj{F}{0,l} w \, v = \int_F w \, v \quad \forall v \in \Pk{l}(F).
\end{align}
These projectors can also be computed locally via a small system involving cell or facet integrals. We also define the so-called difference operators $\delta_T^k : \Uh^{k,l}(T) \to \Pk{k}(T)$ and $\delta_{TF}^l : \Uh^{k,l}(T) \to \Pk{l}(F)$ by
\begin{align*}
  \delta_T^k \ul{v}_T = \lproj{T}{0,k} (\rec \ul{v}_T - v_T), \quad \text{and} \quad
  \delta_{TF}^l \ul{v}_T = \lproj{F}{0,l} (\rec \ul{v}_T - v_F). \nonumber
\end{align*}
Note that local operators, like the weak gradient in the \ac{wg} methods, can also be readily computed in this framework. With these local operators in hand, we can now formulate the discrete problem and describe the assembly process.

\subsection{Cell-local assembly and static condensation} \label{subsec:mot-assembly}

The global discrete problem is assembled from cell-local contributions, which is a third defining characteristic of hybrid methods. Denoting by $\Fh^{\boundary}$ the set of boundary facets, the \ac{hho} approximation of \eqref{eq:poisson-weak} reads: find $\uh \in \underline{U}^{k,l}_{h,g} \doteq  \{ \ul{u}_h \in \Uh^{k,l}: u_F = \lproj{F}{0,l}(g) \ \forall F \in \Fh^{\boundary} \}$ such that
\begin{align}\label{eq:poisson-hybrid-general}
  &\hspace{2.5cm}\qquad a_h(\uh,\vh) = l_h(\vh) \quad \forall \vh \in \underline{U}^{k,l}_{h,0} \doteq \{ \ul{v}_h \in \Uh^{k,l}: v_F = 0 \ \forall F \in \Fh^{\boundary} \},\\
  &\text{where} \quad a_h(\uh,\vh) = \sum_{T \in \Th} \left[ a_T(\uT,\vT) + S_T(\uT,\vT) \right], \quad
  l_h(\vh) = \sum_{T \in \Th} l_T(\vT), \nonumber
\end{align}
with $\uT = \uh|_T$ denoting the restriction of $\uh$ to cell $T$ and its surrounding facets. The local bilinear form $a_T(\uT,\vT)$ represents the consistency term, $S_T(\uT,\vT)$ provides stabilisation and $l_T(\vT)$ is the local load term. The involved local forms are defined as
\begin{align} \label{eq:hho-poisson-forms}
  \begin{aligned}
    &\hspace{2.5cm}\qquad a_T(\uT,\vT)
    = \int_T \nabla \rec \ul{u}_T \cdot \nabla \rec \ul{v}_T , \\
    &S_T(\uT,\vT) = h_T^{-1} \facesum \int_F (\delta_{TF}^l - \delta_{T}^k) \uT \cdot (\delta_{TF}^l - \delta_{T}^k) \vT , \quad \text{and} \quad 
    l_T(\vT) = \int_T f v_T. 
  \end{aligned}
\end{align}
where $h_T$ denotes the diameter of cell $T$.
In the facet integrals defining $S_T$, the cell polynomial $\delta_T^k \ul{v}_T \in \Pk{k}(T)$ is understood as restricted to $F$.
The evaluation of these forms requires integrating over the cell $T$ and its facets, producing a local matrix and vector for each cell that couples the cell unknowns $u_T$ with the facet unknowns $(u_F)_{F \in \Fh(T)}$.

The computational efficiency of hybrid methods stems from \emph{static condensation}: since the cell unknowns appear only in their respective local contributions, they can be eliminated locally to obtain a reduced global system involving only the facet unknowns. This condensation process involves:
\begin{enumerate}
  \item For each cell $T$, assembling the local system coupling cell and facet \acp{dof};
  \item Solving locally for the cell unknowns in terms of the facet unknowns (a Schur complement operation);
  \item Assembling a reduced global system on the mesh skeleton $\Fh$ only.
\end{enumerate}
The reduced system is typically much smaller than the full system, especially for high polynomial degrees, and once solved, the cell unknowns can be recovered efficiently via local solves.

\subsection{Towards a unified computational framework} \label{subsec:mot-framework}

The preceding exposition of \ac{hho} for the Poisson problem reveals a computational structure that is shared across the broader family of hybrid methods (\ac{hdg}, \ac{hho}, \ac{wg}). From an implementation perspective, these methods require the same fundamental building blocks:

\begin{enumerate}
  \item \textbf{Hybrid discrete spaces:} \acp{dof} distributed across mesh entities of different dimensions---typically cells and facets---forming a product space structure. The implementation must handle data structures that associate unknowns with both volumetric elements and the mesh skeleton.

  \item \textbf{Local integration on cells and facets:} All methods require the computation of integrals over each cell $T$ and its boundary facets $F \in \Fh(T)$. These integrals appear in the definition of local operators (e.g., reconstructions, projections) and in the evaluation of local bilinear and linear forms. The implementation must provide efficient quadrature rules and basis function evaluations on general polytopes.

  \item \textbf{Local operator assembly and solution:} Many hybrid methods define auxiliary operators (reconstructions, projections, stabilisations) that are computed by assembling and solving small local linear systems on each cell. The framework must provide mechanisms to assemble these systems from user-defined variational forms and solve them efficiently.

  \item \textbf{Cell-local assembly and static condensation:} The global discrete system is assembled by iterating over cells, computing local contributions that couple cell and facet unknowns, and then condensing out the cell unknowns to form a reduced global system on the skeleton. The implementation must automate this condensation process and manage the assembly of the reduced system.
\end{enumerate}

Despite the diversity of hybrid methods, the computational workflow is remarkably uniform: for each cell, integrate local forms, solve local systems to compute auxiliary operators, assemble local matrices and vectors, perform static condensation and contribute to the global skeleton system. This uniformity suggests that a \emph{generic computational framework} can be designed to support the implementation of a wide range of hybrid methods without requiring low-level, method-specific code for each variant.

The goal of this work is to present such a framework, built on top of the Gridap \ac{fe} library, that abstracts the common computational patterns while remaining flexible enough to accommodate the specific features of different hybrid methods. By providing high-level abstractions for hybrid spaces, local operators and static condensation, the framework enables users to implement complex hybrid discretisations with minimal effort, focusing on the mathematical formulation rather than low-level implementation details. We remark that standard \ac{dg} methods are also accommodated as a special case: by omitting facet unknowns, local reconstruction operators and static condensation, the framework reduces to a conventional \ac{dg} assembly. In the following sections, we describe the design and implementation of this framework and demonstrate its application to several hybrid methods.

\section{Introduction to Gridap.jl} \label{sec:gridap}

Gridap \cite{Gridap,Verdugo2022} is an open-source \ac{fe} framework written entirely in the Julia programming language \cite{Julia} for the numerical approximation of \acp{pde}. Being written exclusively in Julia, Gridap sidesteps the two-language problem common in scientific computing---where performance-critical code is written in C++ or Fortran while user interfaces are provided in Python---by leveraging Julia's \ac{jit} compilation to generate efficient, problem-specific machine code at runtime. This allows researchers to extend the library and develop new numerical methods entirely in a high-level language, without sacrificing performance.

Unlike \ac{fe} libraries that follow traditional object-oriented designs (e.g., Deal.II \cite{dealII93}) or rely on domain-specific form compilers (e.g., FEniCS \cite{fenics}), Gridap adopts a software architecture based on Julia's multiple dispatch paradigm and lazy evaluation. Rather than explicitly iterating over mesh cells, users build objects representing elemental quantities (such as stiffness matrices or load vectors) for all cells simultaneously. These objects are typically \emph{lazy}: the underlying data is computed on-the-fly when needed rather than stored in memory for all cells at once, reducing memory requirements. This design hides assembly loops from user code, yielding a compact syntax that closely resembles mathematical weak forms---a standard Poisson or Stokes problem can be solved in 10--20 lines of code.

Gridap also provides a modular low-level interface based on a small number of fundamental abstractions that stem from mathematical concepts: arrays (\jl{AbstractArray}), callable objects (\jl{Map}), physical fields (\jl{Field}), tensor values (\jl{MultiValue}) and linear functionals (\jl{Dof}). These abstractions can be combined and extended via multiple dispatch, allowing advanced users to customise the library's behaviour without modifying its core. A comprehensive explanation of these software abstractions is provided in \cite{Verdugo2022}, and tutorials can be found in \cite{GridapTutorials}.

\section{New features for hybrid polytopal methods} \label{sec:implementation}

This section presents the new abstractions added to Gridap to support hybrid polytopal methods. We begin by describing how general polytopes are represented using graph structures (\secref{subsec:polytopes}). We then introduce polytopal meshes and their topological queries (\secref{subsec:polytopal-meshes}). Next, we present broken polynomial spaces on arbitrary mesh slices (\secref{subsec:broken-spaces}). We then describe the framework for local assembly and solves (\secref{subsec:local-assembly}), the local operator abstraction (\secref{subsec:local-operators}), and the static condensation procedure (\secref{subsec:static-condensation}).
All features presented here are available in Gridap \texttt{v0.19.8} and later.

\subsection{General representation of polytopes} \label{subsec:polytopes}

We follow the framework used in \cite{BadiaMartorell2022} to represent general polytopes using rotation systems — a subset of planar graphs in which the neighbours of each vertex are arranged in a cyclic order. These graph-based representations enable efficient storage, flexible topological queries and straightforward integration with our \ac{fe} assembly pipeline. For the sake of completeness, we reproduce here the main results.

For a given graph $G$, the vertex set of $G$ is denoted as $\vertices{G}$ and its adjacencies are represented by $\adj{G}$. The set of vertices adjacent to a vertex $\alpha \in \vertices{G}$ is denoted by $\adj{G}(\alpha)$. In a rotation system $R$, for a vertex $\alpha \in \vertices{R}$ and an adjacent vertex $\beta \in \adj{R}(\alpha)$, there exists a clearly defined preceding and succeeding vertex within $\adj{R}(\alpha)$ due to the cyclic ordering. Consequently, we can define $\nextvertex{\alpha}{\beta}$ as the vertex that follows $\beta$ in the cyclic ordering $\adj{R}(\alpha)$. 

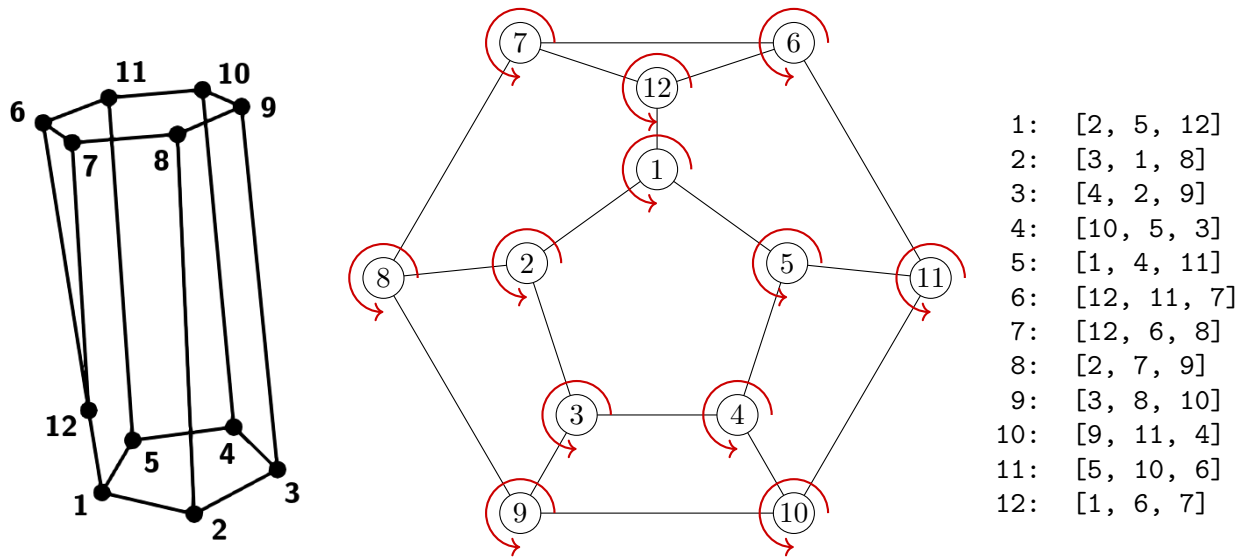
\begin{figure}
  \includegraphics[width=0.27\textwidth]{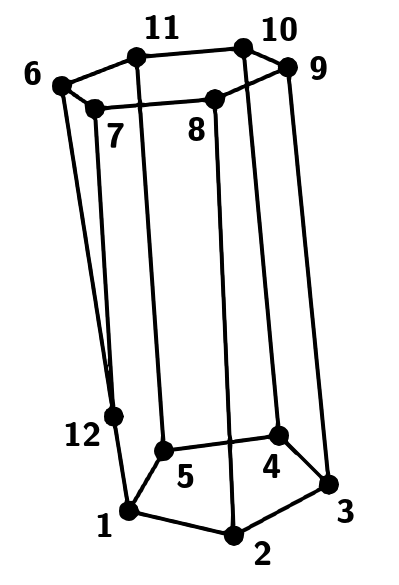}
  \begin{tikzpicture}[x=1.8cm,y=1.8cm]
  \draw [ ] (0.0,1.0) -- (0.0,1.6);
  \draw [ ] (0.0,1.6) -- (0.9999999999999998,1.9320508075688774);
  \draw [ ] (0.0,1.6) -- (-1.0,1.9320508075688774);
  \draw [ ] (-0.9510565162951535,0.3090169943749475) -- (-2.0,0.20000000000000007);
  \draw [ ] (-0.5877852522924732,-0.8090169943749473) -- (-1.0,-1.5320508075688772); 
  \draw [ ] (0.587785252292473,-0.8090169943749475) -- (0.9999999999999994,-1.532050807568878); 
  \draw [ ] (0.9510565162951536,0.3090169943749472) -- (2.0,0.20000000000000062); 

  \draw [ ] (0.0,1.0) -- (-0.9510565162951535,0.3090169943749475); 
  \draw [ ] (-0.9510565162951535,0.3090169943749475) -- (-0.5877852522924732,-0.8090169943749473); 
  \draw [ ] (-0.5877852522924732,-0.8090169943749473) -- (0.587785252292473,-0.8090169943749475); 
  \draw [ ] (0.587785252292473,-0.8090169943749475) -- (0.9510565162951536,0.3090169943749472); 
  \draw [ ] (0.9510565162951536,0.3090169943749472) -- (0.0,1.0); 

  \draw [ ](0.9999999999999998,1.9320508075688774) -- (-1.0,1.9320508075688774); 
  \draw [ ](-1.0,1.9320508075688774) -- (-2.0,0.20000000000000007); 
  \draw [ ](-2.0,0.20000000000000007) -- (-1.0,-1.5320508075688772); 
  \draw [ ](-1.0,-1.5320508075688772) -- (0.9999999999999994,-1.532050807568878); 
  \draw [ ](0.9999999999999994,-1.532050807568878) -- (2.0,0.20000000000000062); 
  \draw [ ](2.0,0.20000000000000062) -- (0.9999999999999998,1.9320508075688774); 

  \draw [fill=white,draw=black] (0.0,1.0) circle (0.15) node[black] {1}; 
  \draw [fill=white,draw=black] (-0.9510565162951535,0.3090169943749475) circle (0.15) node[black] {2};
  \draw [fill=white,draw=black] (-0.5877852522924732,-0.8090169943749473) circle (0.15) node[black] {3};
  \draw [fill=white,draw=black] (0.587785252292473,-0.8090169943749475) circle (0.15) node[black] {4};
  \draw [fill=white,draw=black] (0.9510565162951536,0.3090169943749472) circle (0.15) node[black] {5};

  \draw [->,red!80!black,thick] (0.25,1.0) arc (0:270:0.25);
  \draw [->,red!80!black,thick] (-0.7010565162951535,0.3090169943749475) arc (0:270:0.25); 
  \draw [->,red!80!black,thick] (-0.3377852522924732,-0.8090169943749473) arc (0:270:0.25); 
  \draw [->,red!80!black,thick] (0.837785252292473,-0.8090169943749475) arc (0:270:0.25); 
  \draw [->,red!80!black,thick] (1.2010565162951536,0.3090169943749472) arc (0:270:0.25); 

  \draw [fill=white,draw=black] (0.9999999999999998,1.9320508075688774) circle (0.15) node[black] {6}; 
  \draw [fill=white,draw=black] (-1.0,1.9320508075688774) circle (0.15) node[black] {7}; 
  \draw [fill=white,draw=black] (-2.0,0.20000000000000007) circle (0.15) node[black] {8}; 
  \draw [fill=white,draw=black] (-1.0,-1.5320508075688772) circle (0.15) node[black] {9}; 
  \draw [fill=white,draw=black] (0.9999999999999994,-1.532050807568878) circle (0.15) node[black] {10}; 
  \draw [fill=white,draw=black] (2.0,0.20000000000000062) circle (0.15) node[black] {11}; 

  \draw [->,red!80!black,thick] (1.25,1.9320508075688774) arc (0:270:0.25);
  \draw [->,red!80!black,thick] (-0.75,1.9320508075688774) arc (0:270:0.25);
  \draw [->,red!80!black,thick] (-1.75,0.20000000000000007) arc (0:270:0.25);
  \draw [->,red!80!black,thick] (-0.75,-1.5320508075688772) arc (0:270:0.25);
  \draw [->,red!80!black,thick] (1.25,-1.532050807568878) arc (0:270:0.25);
  \draw [->,red!80!black,thick] (2.25,0.20000000000000062) arc (0:270:0.25);

  \draw [fill=white,draw=black] (0.0,1.6) circle (0.15) node[black] {12};
  \draw [->,red!80!black,thick] (0.25,1.6) arc (0:270:0.25);
\end{tikzpicture}
  \begin{tikzpicture}
  \node[font=\ttfamily, align=left, right=2cm] {
    \ 1: [2, 5, 12] \\
    \ 2: [3, 1, 8] \\
    \ 3: [4, 2, 9] \\
    \ 4: [10, 5, 3] \\
    \ 5: [1, 4, 11] \\
    \ 6: [12, 11, 7] \\
    \ 7: [12, 6, 8] \\
    \ 8: [2, 7, 9] \\
    \ 9: [3, 8, 10] \\
    10: [9, 11, 4] \\
    11: [5, 10, 6] \\
    12: [1, 6, 7] \\
  };
\end{tikzpicture}
  \caption{Different representations for a scutoid. Left: 3D polyhedron representation. Middle: Associated rotation system $R$. Right: Vertex adjacency list $\adj{R}$.}
  \label{fig:scutoid}
\end{figure}

A polytope $P \subset \mathbb{R}^d$ is then defined as a rotation system whose vertices are points in $\mathbb{R}^d$. In 2D, each boundary vertex has exactly two neighbors, so the rotation system reduces to a simple closed loop, recovering the classical polygon representation. In 3D, the cyclic ordering encodes the local face structure around each vertex, and the boundary of $P$ forms an oriented closed surface composed of polygons. In \figref{fig:scutoid}, we illustrate all three of these representations for a scutoid, a type of polyhedron recently discovered in living cells that helps pack them efficiently into curved tissues.

\begin{listing}
  \juliacode{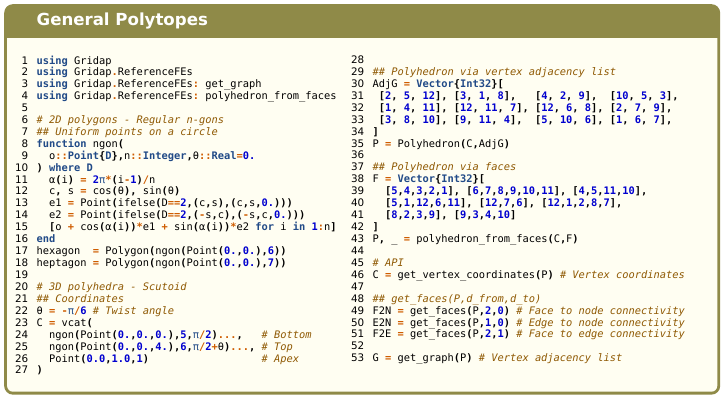}
  \caption{Constructors for general polytopes in Gridap.}
  \label{lst:polytopes}
\end{listing}

\lstref{lst:polytopes} illustrates the construction of general polytopes in Gridap.
In 2D, a \jl{Polygon} is constructed from an ordered list of vertex coordinates (lines 17-18); the cyclic ordering of these vertices implicitly encodes the rotation system, since in 2D the rotation system reduces to a simple closed loop.
In 3D, a \jl{Polyhedron} can be specified in two equivalent ways, both demonstrated using the scutoid depicted in \figref{fig:scutoid}: directly from the vertex adjacency list $\adj{R}$ (lines 29-35), which corresponds precisely to the rotation system $R$ introduced above, or from an enumeration of oriented faces via \jl{polyhedron\_from\_faces} (lines 37-43), from which Gridap infers the rotation system automatically. Once constructed, topological information is extracted via \jl{get\_faces(P, d, s)} (lines 49-51), which returns, for each $d$-face of $P$, the list of its incident $s$-faces, e.g.\ the vertex indices of each 2-face or the edges bounding each face. The underlying rotation system adjacency list $\adj{R}$ can also be recovered directly via \jl{get\_graph(P)} (line 53).

\subsection{Polytopal meshes} \label{subsec:polytopal-meshes}

Traditional simplicial or hexahedral meshes can be efficiently represented by storing a single reference polytope, a list of vertex coordinates and a connectivity table that maps each element to its vertices. With this information, geometrical maps can be computed that map the reference polytope to each of the physical ones. This approach minimises memory usage by avoiding redundant storage of geometric information. 

The same strategy does not work for general polytopal meshes for two fundamental reasons. First, each polytope may be unique, lacking a single reference element. Second, a practical geometrical map from one general polytope to another with the same topology cannot, in general, be defined. Consequently, general polytopal meshes require new data structures that, while more straightforward in design, are potentially more memory-intensive than their traditional counterparts.

\begin{figure}
  \centering
  \includegraphics[width=0.6\textwidth]{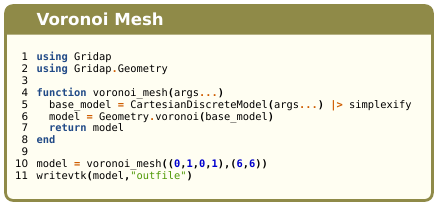}
  \includegraphics[width=0.3\textwidth]{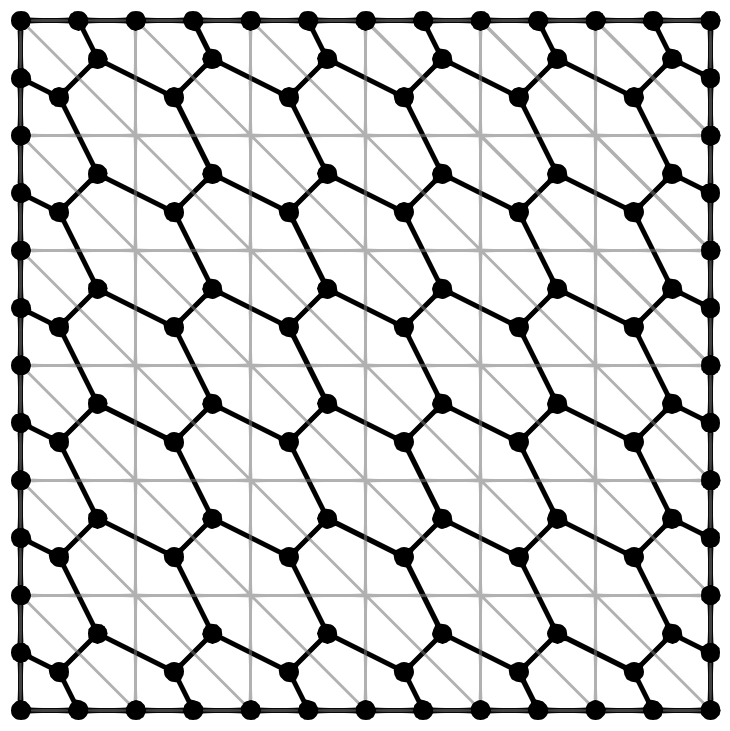}
  \caption{Generation of a 2D Voronoi mesh with Gridap. Left: Julia code to generate the Voronoi mesh. Right: Visualisation of the generated mesh. The initial simplexified Cartesian mesh is shown in light grey for reference.}
  \label{fig:voronoi}
\end{figure}

To this end, we provide a new \jl{PolytopalGridTopology} structure, which stores all the necessary information to represent the topology of a polytopal mesh. This topology is then paired with a \jl{FaceLabeling}, which assigns physical labels to each face of the mesh, to produce a \jl{PolytopalDiscreteModel} object. We currently provide a low-level constructor for \jl{PolytopalDiscreteModel}, as well as several higher-level methods for generating polytopal meshes: Voronoi tessellation from a background simplicial mesh in 2D (illustrated in \figref{fig:voronoi}), and direct conversion of existing simplicial or quadrilateral meshes in both 2D and 3D via the constructor \jl{PolytopalDiscreteModel(model)}. Although not demonstrated in this work, the framework also supports the construction of arbitrarily agglomerated meshes \cite{badia2026gmg}.

We now describe how to extract topological information from a polytopal mesh. Consider a polytopal mesh $\Mh$ of a domain $\dom \subset \mathbb{R}^D$. For any dimension $0 \leq d \leq D$, we refer to $d$-dimensional entities of the mesh as $d$-faces. In particular, $D$-faces correspond to mesh cells (or elements), while $(D-1)$-faces correspond to mesh facets.

We denote by $\Fh^d$ the set of $d$-faces of $\Mh$. In particular, we have that $\Th  \doteq \Fh^D$ and $\Fh \doteq \Fh^{D-1}$. We will use $T$, $F$ and $F^d$ to denote generic elements of $\Th$, $\Fh$ and $\Fh^d$ respectively. 

Given a $d$-face $F^d \in \Fh^{d}$, we denote by $\Fh^{s}(F^d)$ the set of $s$-faces around $F^d$, defined as
\begin{equation} \label{eq:faces_around}
  \Fh^{s}(F^d) \doteq 
  \begin{cases}
    \{ F^s \in \Fh^{s} : F^s \subset \overline{F^d} \} & \text{if } 0 < s < d, \\
    \{ F^d \} & \text{if } s = d \\
    \{ F^{s} \in \Fh^{s} : F^d \subset \overline{F^{s}} \} & \text{if } d < s \leq D.
  \end{cases}
\end{equation}
In particular, the boundary of $F^d$ is given by $\partial F^d \doteq \Fh^{d-1}(F^d)$. For example, the boundary of a cell $T$ consists of its facets: $\partial T = \Fh^{D-1}(T)$. 
In Gridap, this information can be accessed through the \jl{get\_faces(object, d, s)} API, which is defined both at the polytope (see \lstref{lst:polytopes}) and mesh levels. We refer to the provided tutorials \cite{GridapTutorials} for a more detailed explanation. 

\subsection{Broken polynomial spaces} \label{subsec:broken-spaces}

Having established how to represent and query polytopal mesh topology, we now introduce the \ac{fe} spaces defined on these meshes.
All discontinuous and hybrid non-conforming methods rely on broken polynomial spaces defined on the $d$-dimensional faces of the mesh. Within the hybrid methods presented in this paper, we will only be dealing with polynomial spaces on cells ($d = D$) and faces ($d = D-1$). The implementation does, however, support the use of this machinery on faces of arbitrary dimension, opening the door to more complex constructions such as the conforming \ac{vem} \cite{Veiga-Brezzi-Marini-Russo-2023-VEM} or the discrete de Rham sequence on polytopal meshes \cite{discrete_derham}.

We will begin by introducing local polynomial spaces. We denote by $\Pk{k}$ the space of polynomials of degree at most $k$ on $\mathbb{R}^d$. Let $X \subset \mathbb{R}^d$, $d \geq 1$, be an open bounded connected set. The local polynomial space $\Pk{k}(X)$ is defined as the restriction of $\Pk{k}$ to $X$. Given $\mathcal{X}_h \subset \Fh^d$, we can define a broken polynomial space on $\mathcal{X}_h$ as the set of functions that are polynomials on each $d$-face $F^d \in \mathcal{X}_h$, namely
\begin{equation}
  \Pk{k}(\mathcal{X}_h) = \bigtimes_{F^d \in \mathcal{X}_h} \Pk{k}(F^d).
\end{equation}
We provide a new \jl{PolytopalFESpace} type to construct any kind of broken polynomial space on slices of a polytopal mesh, as illustrated in \lstref{lst:polytopal_spaces}. Lines 10--11 first create the cell and facet triangulations $\Omega \doteq \Th$ and $\Gamma \doteq \Fh$ from the discrete model. The vector-valued broken cell space $[\Pk{k}(\Th)]^D$ is then constructed as \jl{V} (lines 14--16) by supplying the triangulation, the value type and the polynomial order $k$.
\begin{listing}[ht!]
  \juliacode{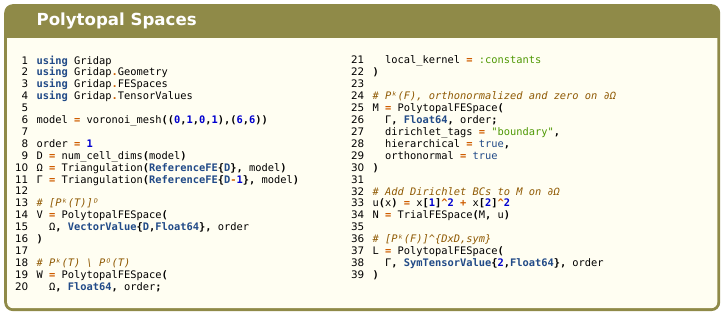}
  \caption{Construction of broken polynomial \ac{fe} spaces on polytopal meshes using the \jl{PolytopalFESpace} constructor. The example shows how to create spaces on different mesh entities (cells, facets) with varying value types and polynomial degrees.}
  \label{lst:polytopal_spaces}
\end{listing}
This construction can be extended to any local polynomial subspace of $\Pk{k}(F^d)$. Namely, for $\mathbb{X}(F^d) \subset \Pk{k}(F^d)$, we define its associated broken polynomial space as
\begin{equation}
  \mathbb{X}(\mathcal{X}_h) = \bigtimes_{F^d \in \mathcal{X}_h} \mathbb{X}(F^d).
\end{equation}
Examples of such spaces will often arise in the literature, such as the space of locally zero-mean polynomials of degree at most $k$, defined on each $d$-face $F^d \in \mathcal{X}_h$ as
\begin{equation}
  \Pk{k}_0(F^d) = \left\{ p \in \Pk{k}(F^d) : \int_{F^d} p \, d\bs{x} = 0 \right\}.
\end{equation}
This zero-mean subspace is constructed by passing \jl{local\_kernel = :constants} to \jl{PolytopalFESpace}, as shown for \jl{W} (lines 19--22), which projects out the locally constant component from each local basis. We will be using this feature in \secref{subsec:example-stokes} to define locally zero-mean pressure spaces for the Stokes problem.

Broken spaces can equally be defined on the facet mesh $\Gamma \doteq \Fh$. Lines 25--30 construct a scalar facet space $\Pk{k}(\Fh)$ as \jl{M}, with Dirichlet boundary conditions imposed on $\partial\dom$ via \jl{dirichlet\_tags} and (optionally) makes it both locally hierarchical and orthonormal. These two properties are often desirable for skeleton spaces in \ac{hho} to maintain convergence rates in the presence of small faces \cite{Badia2022SmallCuts}. The corresponding trial space with prescribed Dirichlet data $u$ is then obtained via the standard \jl{TrialFESpace} constructor (lines 33--34). Finally, \jl{L} (lines 37--39) demonstrates that \jl{PolytopalFESpace} accommodates arbitrary value types by constructing a symmetric tensor-valued facet space $[\Pk{k}(\Fh)]^{D\times D,\text{sym}}$.

These broken polynomial spaces serve as trial and test spaces for hybrid methods, but also as reconstruction targets for local operators, which we describe next.

\subsection{Local assembly and solves} \label{subsec:local-assembly}

Assembling and solving local (usually cell-wise) linear systems is at the heart of hybrid methods. These operations are used, for instance, to statically condense the cell \acp{dof} or to compute local projections such as the \ac{hho} reconstruction operator. As part of this library expansion, we provide a high-level flexible framework to integrate, assemble and solve local \ac{fe} problems defined on subsets of mesh geometrical entities, called {\it patches}. 

Although all the local solves needed in this work are defined on a single cell and its surrounding facets, we designed our code to support arbitrary patch definitions. The framework can, for instance, accommodate multi-cell patches, interface patches between cells sharing multiple faces and overlapping patches where a mesh entity belongs to multiple patches. This flexibility enables applications beyond hybrid methods, such as patch-based smoothers for geometric multigrid solvers, similar to the PCPATCH framework developed in \cite{10.1145/3445791}. Examples of this framework being used within multigrid solvers can be found as part of the GridapSolvers package \cite{GridapSolvers}. However, such applications are beyond the scope of this work.

Our framework revolves around three new objects. First, a \jl{PatchTopology} that defines, for each local problem to be solved, the set of mesh entities (patch) whose contributions have to be considered. Second, \jl{PatchTriangulation} objects that take $D$-dimensional slices of these patches where the different contributions can be defined and integrated. Finally, a \jl{PatchAssembler} that gathers all the contributions from different \jl{PatchTriangulation} objects and assembles them into an array of local matrices, vectors or matrix-vector pairs (local linear systems).

A \jl{PatchTopology} is defined by providing an underlying \jl{GridTopology} object and, for each dimension, a list of the $d$-faces that belong to each patch. For instance, in \lstref{lst:patch_assembly} (line 29), we define patches around each cell $T$ of a polytopal mesh $\Mh$, where each patch comprises the cell $T$ itself and all its surrounding faces, i.e., $\{ T , ( \Fh^d(T) )_{d < D} \}$.

Integration in Gridap revolves around \jl{Triangulation} objects, which represent single-dimensional slices of a mesh (i.e., collections of $d$-faces for a fixed $d$) where integral contributions are defined. We provide two constructors for \jl{Triangulation} objects on patches: a \jl{PatchTriangulation} that contains the $D$-faces for each patch and a \jl{PatchBoundaryTriangulation} that contains, for each patch, the $(D-1)$-faces lying on the patch boundary. Internally, the faces are concatenated for efficient integration, but enough information is retained to facilitate patch-wise integration. \lstref{lst:patch_assembly} (lines 30-31) illustrates how to create both types of \jl{Triangulation} objects for the patches defined above, which will be used to define the constrained local \ac{rhs} and \ac{lhs} for the \ac{hho} reconstruction operator \eqref{eq:hho-reconstruction} (lines 35-40).

Finally, we provide a \jl{PatchAssembler} object that takes care of assembling local contributions defined on the different \jl{PatchTriangulation} objects into local arrays. It supports both single-variable and multi-variable problems, and, in the latter case, it allows local block assembly. \lstref{lst:patch_assembly} (lines 43-47) shows how to use the \jl{PatchAssembler} to assemble the local systems defined in \eqref{eq:hho-reconstruction} (lines 35-40), where the constraint is imposed via Lagrange multipliers. For each patch \jl{p}, \jl{cell\_lhs[p]} and \jl{cell\_rhs[p]} contain the local system matrices, split into $2\times2$ blocks corresponding to the different test and trial variables. Using Gridap's \jl{LazyArray}, we can delay all computations until the local systems are required so these arrays are never explicitly stored in full. This framework also caches and reuses all the memory required to compute and assemble a single local problem, reducing the memory footprint and improving performance. 

Paired with the local assembly of linear systems, we provide several local maps to solve these systems. The most basic is the \jl{LocalSolveMap}, which takes a single monolithic right-hand-side matrix and arbitrarily many left-hand-side arrays, returning the solution of each associated local linear system by means of Julia's standard LU solver. However, implementing more specialised local solvers that exploit the structure of the local systems is straightforward. In \lstref{lst:patch_assembly} (lines 50-74) we show how to implement a custom local solver for the \ac{hho} reconstruction operator, which exploits the $2\times2$ block structure of the local systems to perform a constrained solve. In our implementation, this is done using an exact penalty-like approach, detailed in \cite[Appendix B.2.1]{DiPietro-Droniou-2020-HHO-Book}, where the Lagrange multiplier is eliminated before solving. This custom solver is defined as a \jl{Map}, and its \jl{evaluate!} function will be called for each patch. Thus, the inputs \jl{lhs} and \jl{rhs} correspond to a single pair of local system matrices, and the output is a pair of matrices containing the coefficients of the reconstruction on that patch. Although ignored in the example implementation, the \jl{return\_cache} function can be used to pre-allocate variables that will be reused across different evaluations of \jl{evaluate!}. These caches are then passed as the first argument of \jl{evaluate!}. A detailed explanation of Gridap's \jl{Map} interface can be found in \cite{Verdugo2022}.

In \lstref{lst:patch_assembly}, the arrays of \ac{rhs} and \ac{lhs} matrices lazily assembled in lines 43-47 are fed to the constrained local solver in lines 77-81; then the coefficients of the reconstruction are linearly combined with the local basis of the reconstruction space \jl{L} to generate the reconstructed bases on each cell (lines 84-94). 

\begin{listing}[ht!]
  \juliacode{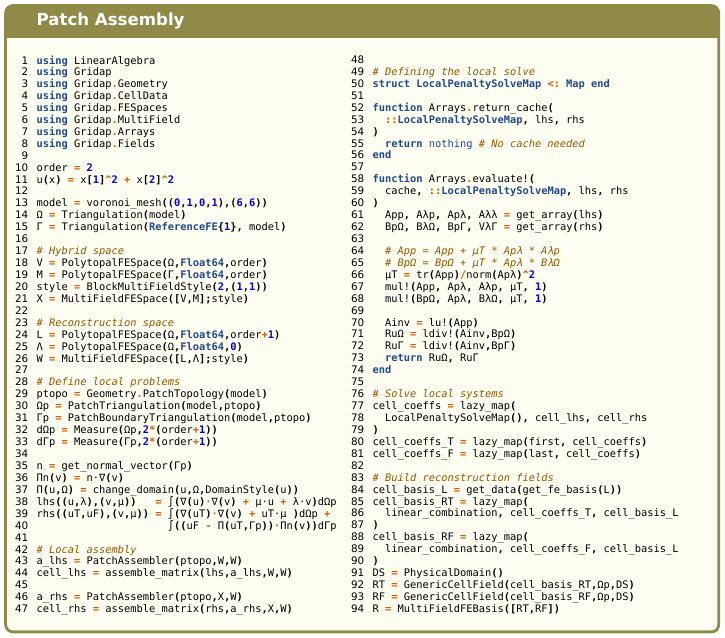}
  \caption{Low-level patch assembly example, showing a low-level implementation of the \ac{hho} reconstruction operator defined in \eqref{eq:hho-reconstruction}.}
  \label{lst:patch_assembly}
\end{listing}

\subsection{Local operators} \label{subsec:local-operators}

Local operators encapsulate the common pattern of computing discrete projections through local linear solves. Like in the example showcased in \lstref{lst:patch_assembly}, the solutions of the local linear systems are usually used to define projections between two different finite dimensional spaces. For instance, the \ac{hho} reconstruction operator defines a discrete elliptic projector from the local \ac{hho} space of order $(k,l)$ into the space of polynomials of degree $k+1$ on the cell. Although these projections can be quite varied, they often share a common structure and therefore most of the low-level complexity shown in \lstref{lst:patch_assembly} can be abstracted away.

To encompass all the possibilities, we provide a new \jl{LocalOperator} structure. In essence, this object takes a recipe to compute the local linear systems, a \jl{Map} to solve each one of these systems and an output discontinuous \jl{FESpace} to which the resulting local solutions belong. 

There are two main signatures, both showcased in \lstref{lst:lop_poisson}. The first one is reserved for local operators whose linear systems are defined on a single geometrical entity (e.g.\ a single cell or facet). It is used (lines 2--11) to create the $L^2$-projection $\lproj{T}{0,k}$ defined in \eqref{eq:proj_ops}. The \jl{LocalOperator} constructor takes a \jl{LocalSolveMap}, the output \jl{FESpace} and two bilinear forms corresponding to the left- and right-hand sides of the projection, which in this case are both the $L^2$-mass bilinear form.
The second signature is for local operators that require the assembly of linear systems on patches. It is used (lines 13--39) to implement the \ac{hho} reconstruction operator $\rec$ defined in \eqref{eq:hho-reconstruction}, replacing most of the manual assembly shown in \lstref{lst:patch_assembly}. The constructor takes a \jl{LocalPenaltySolveMap}, a \jl{PatchTopology}, the output and input spaces and the bilinear forms \jl{lhs} and \jl{rhs} corresponding to the left- and right-hand sides of~\eqref{eq:hho-reconstruction}. The auxiliary Lagrange multiplier space \jl{$\Lambda$}, used to enforce the zero-mean condition in the second line of~\eqref{eq:hho-reconstruction}, is bundled together with the reconstruction space \jl{L} into a \jl{MultiFieldFESpace} \jl{W} (line 33).

A subtle but important detail is the use of \jl{FESpaceWithoutBCs} (lines 5 and 32).
Within the \ac{hho} context, we use local operators to compute local changes of basis between two \ac{fe} spaces. For instance, the input space to the \ac{hho} reconstruction operator is the local \ac{hho} space, which combines cell \acp{dof} and facet \acp{dof}. The reconstruction operator is a change of basis from this full local \ac{hho} space to the polynomial reconstruction space, so it must act on \emph{all} \acp{dof} within each patch, including facet \acp{dof} on the domain boundary. However, by default, Gridap eliminates Dirichlet-constrained \acp{dof} (i.e.\ boundary facet \acp{dof}) from the assembly process; these would be missing from the columns of the right-hand side matrix assembled for the reconstruction. The same issue would arise for the left-hand-side matrix in the $L^2$-projection operator, where the output space is often the \ac{hho} skeleton space.
The \jl{FESpaceWithoutBCs} wrapper reindexes Dirichlet \acp{dof} as free while preserving the local ordering, ensuring that all \ac{hho} \acp{dof} within each patch (including Dirichlet ones) participate in the local assembly and produce a complete change-of-basis matrix.

%

\begin{listing}[ht!]
  \juliacode{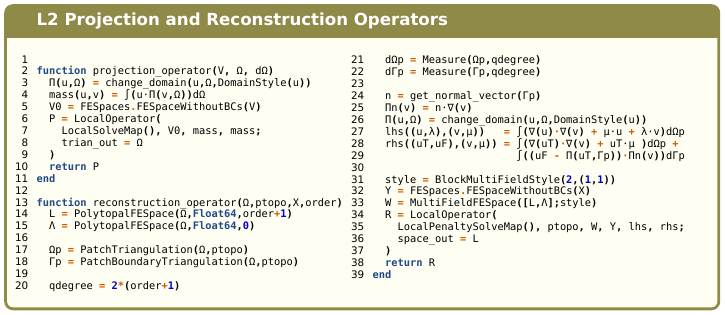}
  \caption{$L^2$-projection and \ac{hho} reconstruction operators, defined using \jl{LocalOperator} on cell patches, abstracting the manual assembly shown in \lstref{lst:patch_assembly}. This high-level interface hides the low-level details of patch assembly and local solves.}
  \label{lst:lop_poisson}
\end{listing}

\subsection{Static condensation} \label{subsec:static-condensation}

A defining feature of hybrid methods is the ability to eliminate cell unknowns locally through static condensation, yielding a reduced global system involving only the skeleton (facet) \acp{dof}. This process requires assembling local contributions that couple cell and facet unknowns, then performing a Schur complement operation on each cell to eliminate the cell \acp{dof} before global assembly. Our framework automates this procedure by building on the patch assembly infrastructure introduced in \secref{subsec:local-assembly}.

A key challenge is that local operators such as $\rec$ map from the hybrid space $\Uh^{k,l}(T)$ to a reconstruction space $\Pk{k+1}(T)$ defined on the cell. The resulting reconstructed basis is attached to all hybrid \acp{dof} within the patch, which has to be accounted for during assembly.

To address this, we introduce the \jl{PatchFESpace} construct. Given a space \jl{X} and a \jl{PatchTopology}, the \jl{PatchFESpace}, illustrated in \lstref{lst:static_condensation} (line 57) by \jl{Xp = PatchFESpace(X,ptopo)}, creates a view of \jl{X} that gathers all \acp{dof} in each patch. In the \ac{hho} case, \jl{Xp} gathers, for each cell $T$, the local \acp{dof} from the cell and all its surrounding facets, providing the assembler with the correct global \ac{dof} indices for contributions involving the reconstructed basis.

This leads to a natural $2\times2$ block structure when assembling forms that mix the original and reconstructed bases. On the one hand, the consistency term $a_T$ only involves the reconstruction operator and can therefore be assembled using only \jl{Xp}. On the other hand, the stabilisation term $S_T$ involves both the original and reconstructed bases. This requires that we decompose the facet residual appearing in the stabilisation. Specifically, for any $\ul{u}_T \in \Uh^{k,k}$ and any facet $F \in \Fh(T)$, consider
\[
  s_{TF}(\ul{u}_T) \doteq (\delta_{TF}^k - \delta_T^k)\ul{u}_T,
\]
where $\delta_T^k\ul{u}_T \in \Pk{k}(T)$ is understood as restricted to $F$ in facet expressions. Using the definitions of $\delta_T^k$ and $\delta_{TF}^k$, we can write, on each facet $F$,
\begin{align*}
  s_{TF}(\ul{u}_T) = S_1(\uT) + S_2(\rec \uT),
\end{align*}
with
\begin{align*}
  S_1(\uT) \doteq (u_T)|_F - u_F, \quad \text{and} \quad
  S_2(\rec \uT) \doteq \lproj{F}{0,k}\bigl((\rec \ul{u}_T)|_F\bigr) - \bigl(\lproj{T}{0,k}(\rec \ul{u}_T)\bigr)|_F.
\end{align*}
The terms $S_1$ and $S_2$ are attached to the original and reconstructed \acp{dof}, respectively. The stabilisation can then be rewritten as
\begin{align}
S_T(\uT,\vT) &= h_T^{-1} \facesum \int_{F} \bigl(S_1(\uT) + S_2(\rec \uT)\bigr) \cdot \bigl(S_1(\vT) + S_2(\rec \vT)\bigr)\\
  &= h_T^{-1} \facesum \int_{F} S_1(\uT) \cdot S_1(\vT)
  + h_T^{-1} \facesum \int_{F} S_1(\uT) \cdot S_2(\rec \vT) \\
  &+ h_T^{-1} \facesum \int_{F} S_2(\rec \uT) \cdot S_1(\vT)
  + h_T^{-1} \facesum \int_{F} S_2(\rec \uT) \cdot S_2(\rec \vT)
\end{align}
Lines 66--72 in \lstref{lst:static_condensation} pass these four blocks to \jl{collect\_and\_merge\_cell\_matrix\_and\_vector}, which assembles them into local 2$\times$2 block matrices coupling cell and facet unknowns. Each block is assembled with the appropriate space pair: \jl{(X, X)} for contributions involving only original \acp{dof}, \jl{(Xp, Xp)} for terms involving the reconstruction on both trial and test sides and the mixed pairs \jl{(X, Xp)} and \jl{(Xp, X)} for cross terms. The \jl{PatchAssembler} ensures that contributions are placed at the correct locations in the global matrix based on the \ac{dof} indices provided by each space.

\lstref{lst:static_condensation} demonstrates both the underlying mechanics and the high-level interface. Lines 75--87 show a custom \jl{StaticCondensationMap} that explicitly performs the Schur complement: it extracts the 2$\times$2 matrix blocks $K_{ii}$ (cell-cell), $K_{bb}$ (facet-facet), $K_{ib}$ (cell-facet) and $K_{bi}$ (facet-cell), as well as the \ac{rhs} blocks $b_i$ and $b_b$. It then eliminates the interior \acp{dof} by computing 
\[
  \tilde{K}_{bb} = K_{bb} - K_{bi} K_{ii}^{-1} K_{ib} \quad \text{and} \quad
  \tilde{b}_b = b_b - K_{bi} K_{ii}^{-1} b_i.
\]
Lines 89--97 then manually assemble the condensed skeleton system. For typical use cases, the high-level \jl{StaticCondensationOperator} (lines 100--103) automates this entire process, allowing users to express the method declaratively without managing low-level assembly details.
\begin{listing}[ht!]
  \juliacode{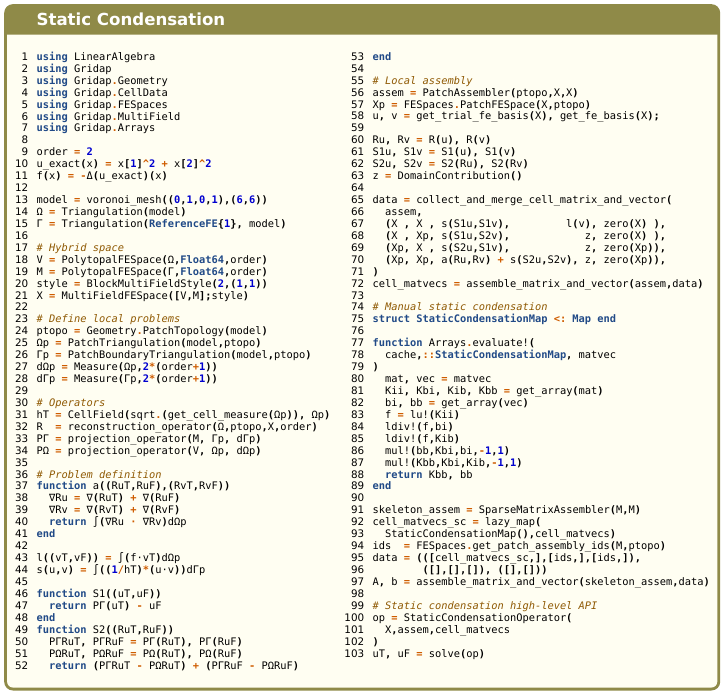}
  \caption{Static condensation example showing the 2$\times$2 block assembly structure. Lines 66--72 demonstrate how contributions are split between original \acp{dof} (X) and patch \acp{dof} (Xp). The listing shows both the low-level mechanics (lines 75--97) and the high-level interface (lines 100--103).}
  \label{lst:static_condensation}
\end{listing}

\subsection{Final driver} \label{subsec:final-driver}

The abstractions presented in this section---polytopal representations, broken polynomial spaces, patch-based assembly, local operators and automated static condensation---form a cohesive framework for implementing hybrid methods. By building on Gridap's lazy evaluation and multiple dispatch, these tools provide both high-level expressiveness and computational efficiency. A complete implementation of a Poisson problem with \ac{hho} discretisation is shown in \lstref{lst:example_hho_poisson}, which uses all the building blocks described in this section. 

In the following section, we demonstrate how to modify these components to enable concise implementations of \ac{dg}, \ac{hdg} and \ac{hho} methods for increasingly complex problems.

\begin{listing}[ht!]
  \juliacode{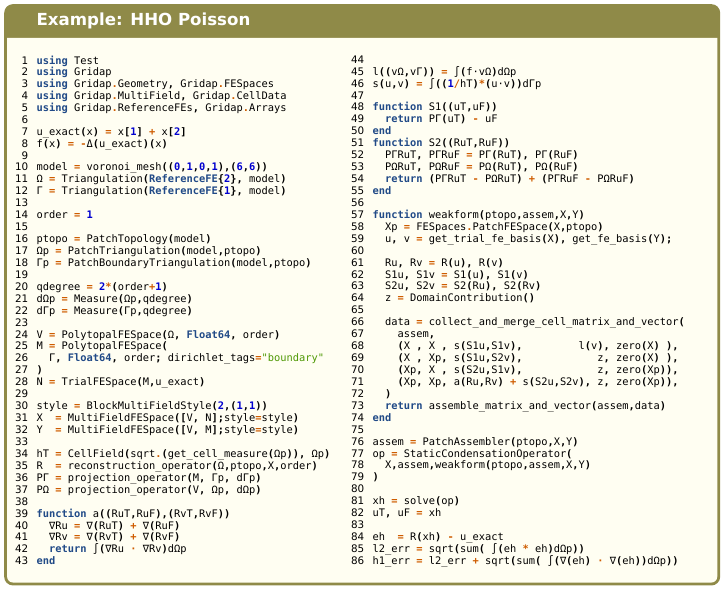}
  \caption{\ac{hho} discretisation of the Poisson problem (\secref{subsec:example-poisson-hho}). Requires {the code in} \figref{fig:voronoi} and {the code in} \lstref{lst:lop_poisson}.}
  \label{lst:example_hho_poisson}
\end{listing}

\section{Examples} \label{sec:examples}

\subsection{A first example: The Poisson problem} \label{subsec:ex-poisson}



As a first example, we consider the Poisson problem \eqref{eq:poisson-weak}.  
In the following subsections, we demonstrate the implementation of three methods of increasing complexity using the framework introduced in \secref{sec:implementation}.
We begin with \ac{dg} to demonstrate basic usage of broken polynomial spaces (\secref{subsec:example-poisson-dg}). Then, we present \ac{hdg}, which introduces hybrid spaces and static condensation (\secref{subsec:example-poisson-hdg}). Finally, we show results for \ac{hho}, which additionally uses local projection operators (\secref{subsec:example-poisson-hho}).

In all three examples, we will use the method of manufactured solutions in a domain $\dom = [0,1]^2$, and consider an exact solution $u^* = x_1^2 + x_2^{2}$. We take $f = -\Delta u^*$ and $g = u^*|_{\partial \dom}$. We will mesh our domain $\dom$ using a Voronoi tessellation $\Mh$ created from a simplexified Cartesian mesh with $n \times n$ cells using the code provided in \figref{fig:voronoi}. 

\subsubsection{\ac{dg} discretisation of the Poisson problem} \label{subsec:example-poisson-dg}

\lstref{lst:example_dg_poisson} shows the implementation of the \ac{dg} discretisation of the Poisson problem \cite{cangiani2014hp}.
Given a polynomial degree $k > 0$, we take our space of unknowns as the broken polynomial space $V_h = \Pk{k}(\Mh)$ (line 13).
Given an interior facet $F \in \Fhi$ and its two adjacent cells $T, T' \in \Th$, we define the jump and average operators for a scalar function $v$ and a tensor function $\bm{\sigma}$ as
\begin{align*}
  \{\!\{ \bm{\sigma} \}\!\} = \frac{1}{2} \left( \bm{\sigma}_{|T} + \bm{\sigma}_{|T'} \right), \quad
  [\![ v ]\!] = v_{|T} \bm{n}_{FT} + v_{|T'} \bm{n}_{FT'},
\end{align*}
where $\bm{n}_{FT}$ and $\bm{n}_{FT'}$ are the unit normal vectors on $F$ pointing outwards from $T$ and $T'$ respectively (lines 24-25). For a boundary facet $F \in \Fhb$ adjacent to a cell $T \in \Th$, we define
\begin{align*}
  \{\!\{ \bm{\sigma} \}\!\} = \bm{\sigma}_{|T}, \quad
  [\![ v ]\!] = v_{|T} \bm{n}_{FT}.
\end{align*}

\begin{figure}[ht!]
  \centering
  \includegraphics[width=0.48\textwidth]{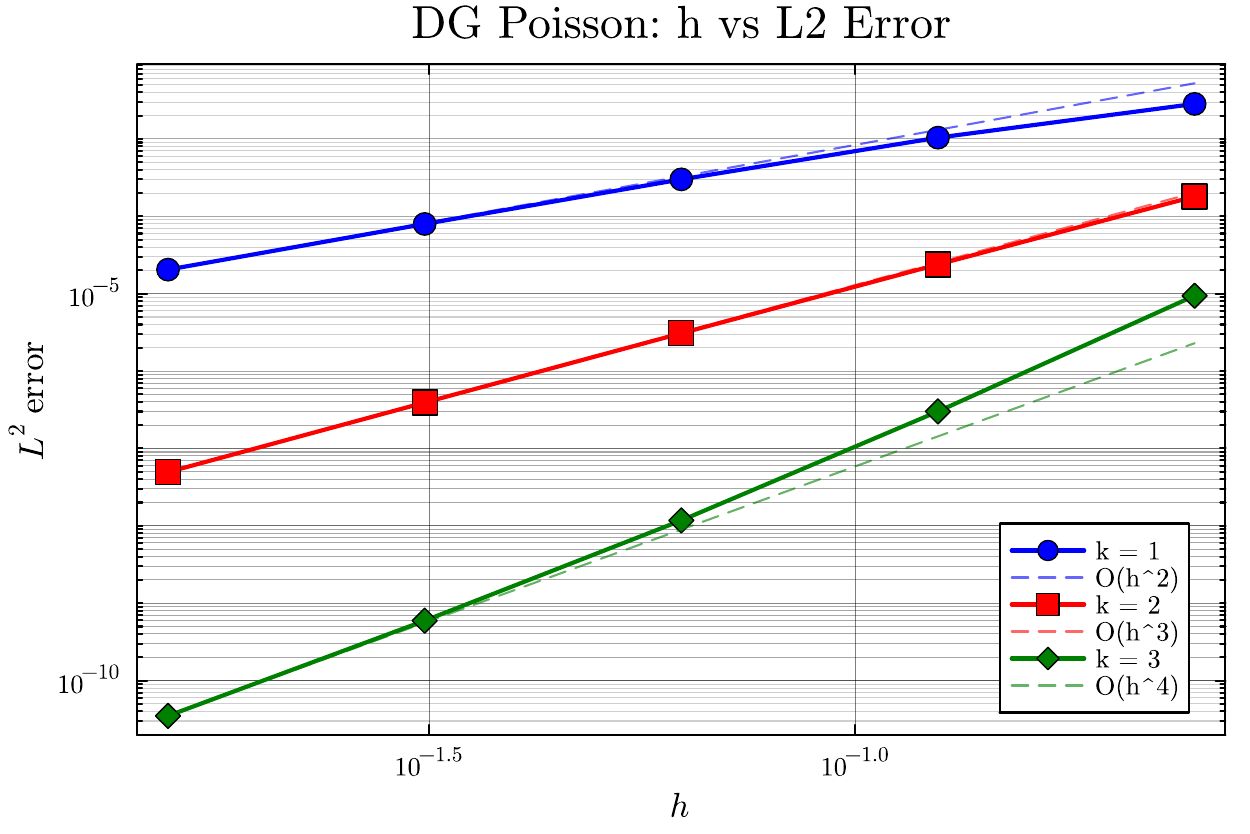}
  \hfill
  \includegraphics[width=0.48\textwidth]{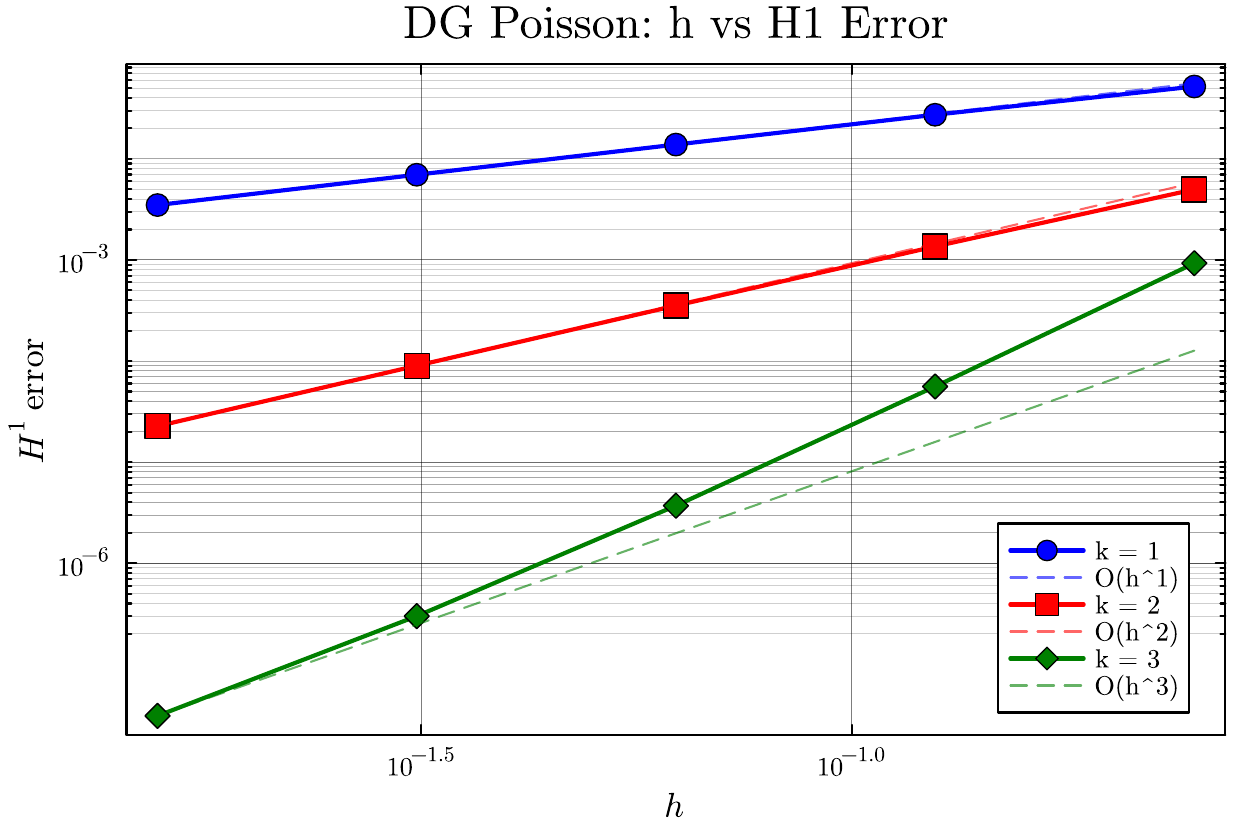}
  \caption{\ac{dg} method convergence for the Poisson problem: $L^2$-error (left) and $H^1$-error (right) versus mesh size $h$ for polynomial degrees $k=1,2,3$. Reference lines show the expected convergence rates in the respective norms. Results generated using the driver from \lstref{lst:example_dg_poisson} on Voronoi meshes constructed as described in \figref{fig:voronoi}.}
  \label{fig:convergence-dg-poisson}
\end{figure}

The discrete problem (lines 30-45) reads: find $u_h \in V_h$ such that 
\begin{equation*}
  a_h(u_h,v_h) = l_h(v_h) \quad \forall v_h \in V_h,
\end{equation*}
where
\begin{align*}
  a_h(u,v) &= \sum_{T \in \Th} \int_T \nabla u \cdot \nabla v \, dx 
         + \sum_{F \in \Fh} \int_{F} \left( \frac{\gamma}{h_{F}} [\![ v ]\!] [\![ u ]\!] - \{\!\{ \nabla u \}\!\} \cdot [\![ v ]\!] - \{\!\{ \nabla v \}\!\} \cdot [\![ u ]\!] \right) ds \\
  l_h(v) &= \sum_{T \in \Th} \int_T f v \, dx + \sum_{F \in \Fhb} \int_{F} \left( \frac{\gamma}{h_{F}} g v - (\nabla v \cdot \bm{n}_{F}) g \right) ds,
\end{align*}
where $\gamma > 0$ is a penalty parameter (line 29) and $h_F$ is the diameter of the facet $F$ (lines 26-27).
\begin{listing}[ht!]
  \juliacode{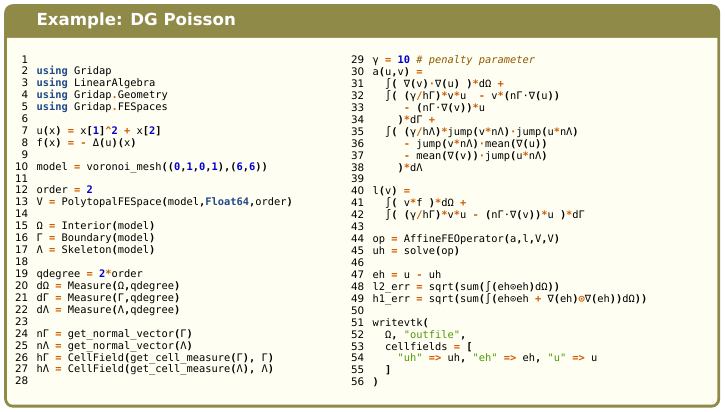}
  \caption{\ac{dg} discretisation of the Poisson problem (\secref{subsec:example-poisson-dg}). Requires {the code in} \figref{fig:voronoi}.}
  \label{lst:example_dg_poisson}
\end{listing}
Convergence results for this method can be found in \figref{fig:convergence-dg-poisson}.


\subsubsection{\ac{hdg} discretisation of the Poisson problem} \label{subsec:example-poisson-hdg}

The \ac{hdg} method \cite{Cockburn_2009} rewrites the Poisson problem as a first-order system by introducing the flux $\bm{p} = -\nabla u$ as an auxiliary variable. The problem then reads: find $(u, \bm{p})$ such that
\begin{align*}
  \bm{p} + \nabla u = \boldsymbol{0} \quad \text{in} \quad \dom, \quad
  \nabla \cdot \bm{p} = f \quad \text{in} \quad \dom, \quad
  u = g \quad \text{on} \quad \partial\dom.
\end{align*}
\lstref{lst:example_hdg_poisson} shows the implementation of this example. Given a polynomial degree $k \geq 0$, we define three spaces of unknowns: The space of cell unknowns $V_h = \Pk{k+1}(\Th)$ (line 17), the space of fluxes $Q_h = \vecPk{k}(\Th)$ (line 18-20) and the space of traces $M_h = \Pk{k}(\Fh)$ (line 21-24). Our hybrid space is then defined as $\Uh = Q_h \times V_h \times M_h$ (lines 26-27), with solutions of the form $\uh  = (\bm{p}_h, u_h, s_h) \in \Uh$. For each cell $T \in \Th$, we denote by $\Uh(T) = Q_h|_T \times V_h|_T \times M_h|_T = \left[\Pk{k}(T)\right]^d \times \Pk{k+1}(T) \times \Pk{k}(\Fh(T))$ the local space of unknowns on $T$, which contains local solutions of the form $\uT = (\bm{p}_T, u_T, s_T) \in \Uh(T)$.
Note that \texttt{BlockMultiFieldStyle(2,(2,1))} (line 26) structures the three-field space as a two-block system $(Q_h \times V_h, M_h)$, where the first block contains all the variables that will be statically condensed.

\begin{figure}[ht!]
  \centering
  \includegraphics[width=0.48\textwidth]{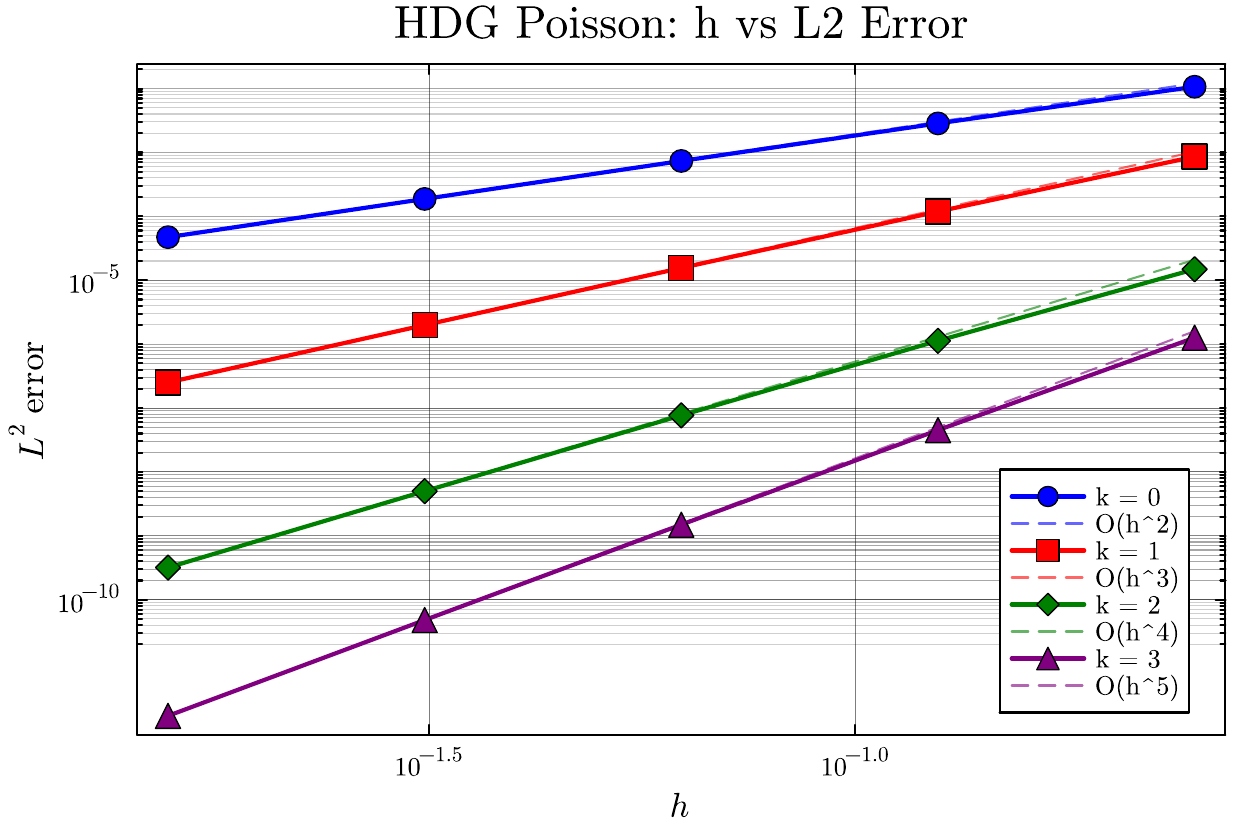}
  \hfill
  \includegraphics[width=0.48\textwidth]{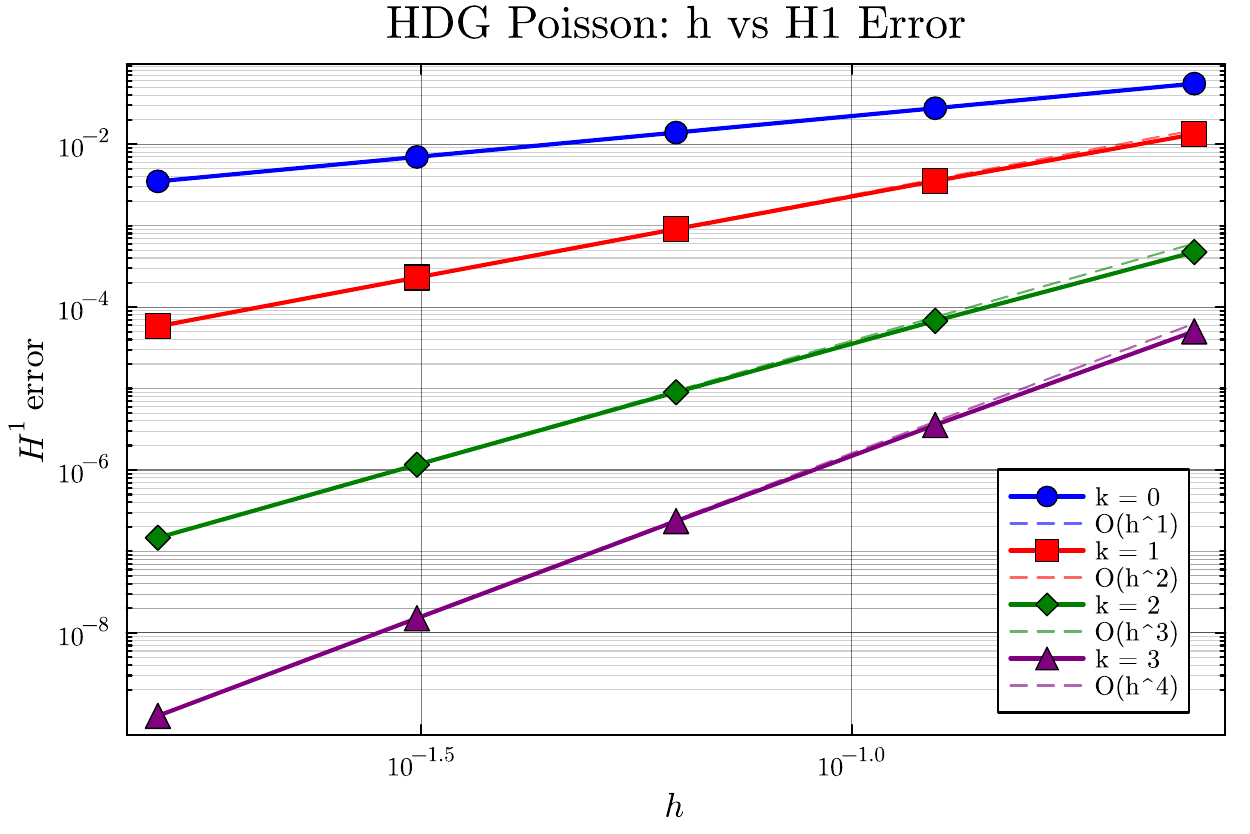}
  \caption{\ac{hdg} method convergence for the Poisson problem: $L^2$ error (left) and $H^1$ error (right) versus mesh size $h$ for polynomial degrees $k=0,1,2,3$. Reference lines show the theoretical convergence rates $O(h^{k+2})$ for $L^2$ and $O(h^{k+1})$ for $H^1$. Results generated using the driver from \lstref{lst:example_hdg_poisson} on Voronoi meshes constructed as described in \figref{fig:voronoi}.}
  \label{fig:convergence-hdg-poisson}
\end{figure}

The discretised problem reads (lines 37-47): find $\uh \in \Uh$ such that $\forall \vh \in \Uh$ we have
\begin{align*}
  &\hspace{3.5cm} a_h(\uh,\vh) = l_h(\vh), \\
  &a_h(\uh,\vh) = \sum_{T \in \Th} a_T(\uT,\vT) + S_T(\ul{u}_T, \ul{v}_T), \quad
  l_h(\vh) = \sum_{T \in \Th} l_T(\vT),
\end{align*}
where
\begin{align*}
  &a_T(\uT,\vT) = \int_T \left( \bm{p}_T \cdot \bm{q}_T - u_T \nabla \cdot \bm{q}_T - \bm{p}_T \cdot \nabla v_T \right)
  + \sum_{F \in \Fh(T)} \int_F \left( s_F \bm{q}_T \cdot \bm{n}_{TF} + \bm{p}_T \cdot \bm{n}_{TF} m_F \right), \\
  &S_T(\uT,\vT) = \tau \sum_{F \in \Fh(T)} \int_F \left((\lproj{F}{0,k} u_T - s_F)(\lproj{F}{0,k} v_T + m_F) \right), \quad  \text{and} \quad
  l_T(\vT) = \int_T f v_T \, dx.
\end{align*}
Note that the stabilisation term $S_T(\uT,\vT)$ is not the standard \ac{hdg} stabilisation, but a slightly modified version that is robust on polytopal meshes \cite{Du_2020}. Convergence results for this method can be found in \figref{fig:convergence-hdg-poisson}.
\begin{listing}[ht!]
  \juliacode{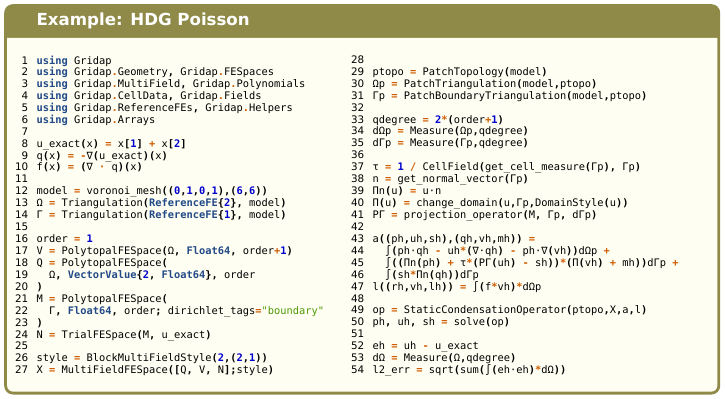}
  \caption{\ac{hdg} discretisation of the Poisson problem (\secref{subsec:example-poisson-hdg}). Requires {the code in} \figref{fig:voronoi} and {the code in} \lstref{lst:lop_poisson}.}
  \label{lst:example_hdg_poisson}
\end{listing}

\subsubsection{\ac{hho} discretisation of the Poisson problem} \label{subsec:example-poisson-hho}

\begin{figure}[ht!]
  \centering
  \includegraphics[width=0.48\textwidth]{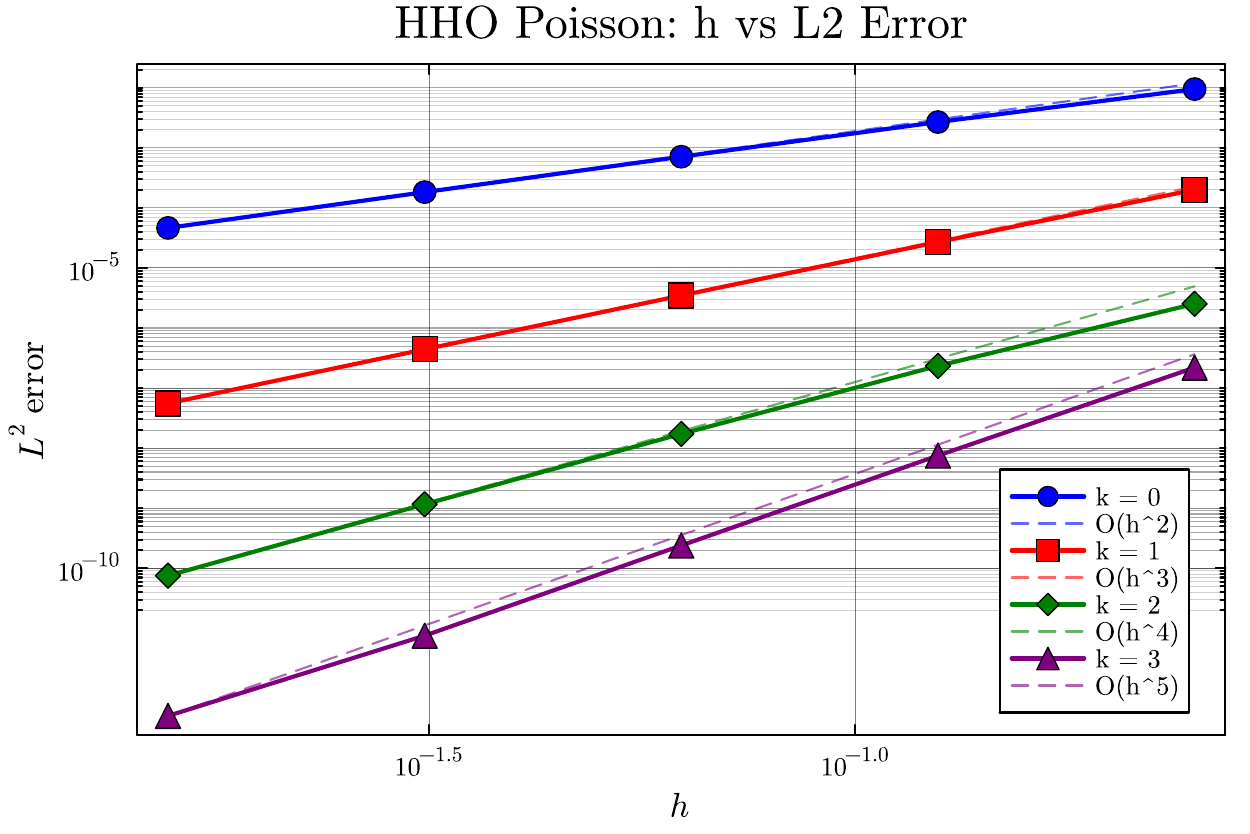}
  \hfill
  \includegraphics[width=0.48\textwidth]{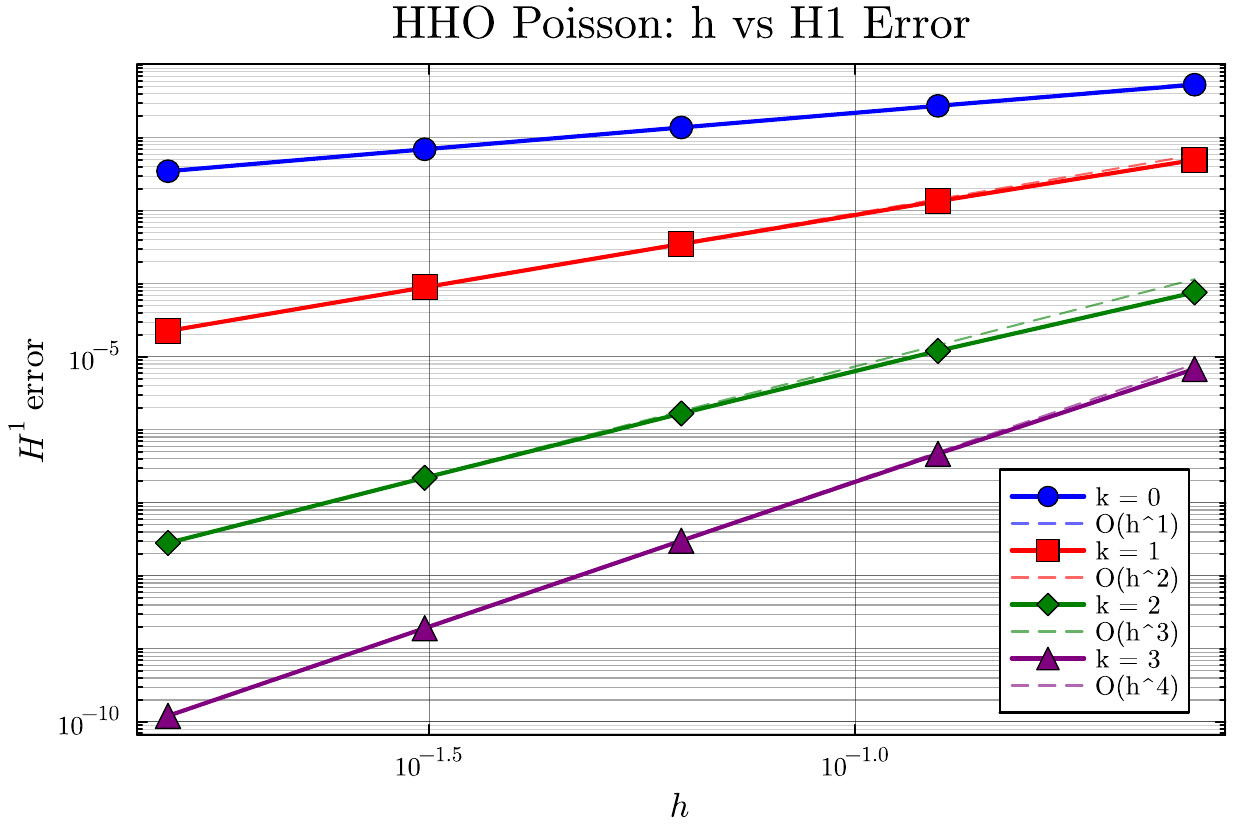}
  \caption{\ac{hho} method convergence for the Poisson problem: $L^2$ error (left) and $H^1$ error (right) versus mesh size $h$ for polynomial degrees $k=0,1,2,3$. Reference lines show the theoretical convergence rates $O(h^{k+2})$ for $L^2$ and $O(h^{k+1})$ for $H^1$, demonstrating superconvergence. Results generated using the driver from \lstref{lst:example_hho_poisson} on Voronoi meshes constructed as described in \figref{fig:voronoi}.}
  \label{fig:convergence-hho-poisson}
\end{figure}

The \ac{hho} method for the Poisson problem was presented in detail in \secref{sec:motivation-poisson}. \lstref{lst:example_hho_poisson} shows the complete implementation of this method using the framework introduced in \secref{sec:implementation}. See \lstref{lst:lop_poisson} for the implementation of the reconstruction operator. Convergence results for this method can be found in \figref{fig:convergence-hho-poisson}. 


\subsection{More involved examples} \label{subsec:ex-advanced}
Now we discuss some more complex standard examples from the literature, showcasing the proposed framework's flexibility and ease of extension to new formulations. 

\subsubsection{\ac{hho} for linear elasticity} \label{subsec:example-elasticity}

Linear elasticity is a vector-valued problem whose 
kernel contains the so-called rigid body modes. The \ac{hho} formulation for linear elasticity \cite{DIPIETRO20151} closely resembles the one for the Poisson problem introduced above, with the main changes being the modification of the local reconstruction operator to account for rigid body motions and the use of vector-valued spaces. We will closely follow the formulation introduced in \cite[Chapter 7]{DiPietro-Droniou-2020-HHO-Book}.  \lstref{lst:example_mhho_elasticity} shows the implementation of this example. To add some variation to the previous example, we will consider the mixed-order \ac{hho} formulation. Given a polynomial degree $k \geq 0$, we define our hybrid space (lines 23-31) as 
\begin{equation} \label{eq:mhho-space-vector}
  \vUh^{k+1,k} = \vecPk{k+1}(\Th) \times \vecPk{k}(\Fh). 
\end{equation}
%
The local reconstruction operator $\symrec : \vUh^{k+1,k}(T) \to \left[\Pk{k+1}(T)\right]^d$ is now uniquely defined by
\begin{align} \label{eq:hho-reconstruction-elasticity}
  (\nabla_s \symrec \ul{v}_T, \nabla_s w)_T &= (\nabla_s v_T, \nabla_s w)_T + \sum_{F \in \Fh(T)} (v_F - v_T , \nabla_s w \cdot \bm{n}_{TF})_F \quad \forall w \in \vecPk{k+1}(T), \\
  (\symrec \ul{v}_T, 1)_T &= (v_T, 1)_T\\
  (\nabla_s \symrec \ul{v}_T, 1)_T &= \sum_{F \in \Fh(T)} \frac{1}{2} (\bm{n}_{TF}\otimes v_F - v_F \otimes \bm{n}_{TF},1)_F
\end{align}
Note that this new reconstruction operator solves a modification of the vector Poisson problem (see \eqref{eq:hho-reconstruction}) in which we replace the gradient by the symmetric gradient $\nabla_s$ and add rigid-body motions to the kernel space. The implementation of this operator can be found in \lstref{lst:lop_elasticity}. Constraints are enforced via Lagrange multipliers, using cell-wise constant vector-valued spaces with one component per constraint: $\Lambda 1$ enforces the mean value (component-wise, $d$ equations), while $\Lambda 2$ enforces the rigid body modes ($d(d+1)/2$ equations) using inlined functions $\alpha$ (component-wise contraction) and $\beta(\bm{n},v) = (1/2)(\bm{n}\otimes v - v \otimes \bm{n})$.

The advantage of mixed-order \ac{hho} for the elasticity problem is that we can replace $\symrec$ by a simpler gradient reconstruction operator $\gradrec : \vUh^{k+1,k}(T) \to \left[\Pk{k+1}(T)\right]^{d \times d}_{\sym}$ mapping to the subspace of symmetric 2-tensors, defined by
\begin{align} \label{eq:hho-gradient-reconstruction-elasticity}
  (\gradrec \ul{v}_T, \bm{\tau})_T &= (\nabla_s v_T, \bm{\tau})_T + \sum_{F \in \Fh(T)} (v_F - v_T , \bm{\tau} \cdot \bm{n}_{TF})_F \quad \forall \bm{\tau} \in \left[\Pk{k+1}(T)\right]^{d \times d}_{\sym}.
\end{align}
The implementation of this operator can also be found in \lstref{lst:lop_elasticity}.

Back in \lstref{lst:example_mhho_elasticity}, the discretised problem then reads (lines 50-72): find $\uh \in \vUh^{k+1,k}$ such that $ a_h(\uh,\vh) = l_h(\vh) \quad \forall \vh \in \vUh^{k+1,k}$. The involved local forms are defined by
\begin{align*}
  &a_T(\uT,\vT)
  = \left( \gradrec \ul{u}_T, \bm{C} : \gradrec \ul{v}_T \right)_T, \\
  &S_T(\uT,\vT) = \sum_{F \in \Fh(T)} h_F^{-1} \int_F \left( (\lproj{F}{0,k}u_T - u_F) \cdot (\lproj{F}{0,k}v_T - v_F) \right), \quad \text{and} \quad
  l_T(\vT) = \int_T f v_T \, dx,
\end{align*}
where $\bm{C}$ is the elasticity tensor, which we take as the isotropic elasticity tensor defined by the Lamé parameters $\lambda$ and $\mu$ (lines 9-12). Convergence results for this method can be found in \figref{fig:convergence-mhho-advanced}.
\begin{listing}[ht!]
  \juliacode{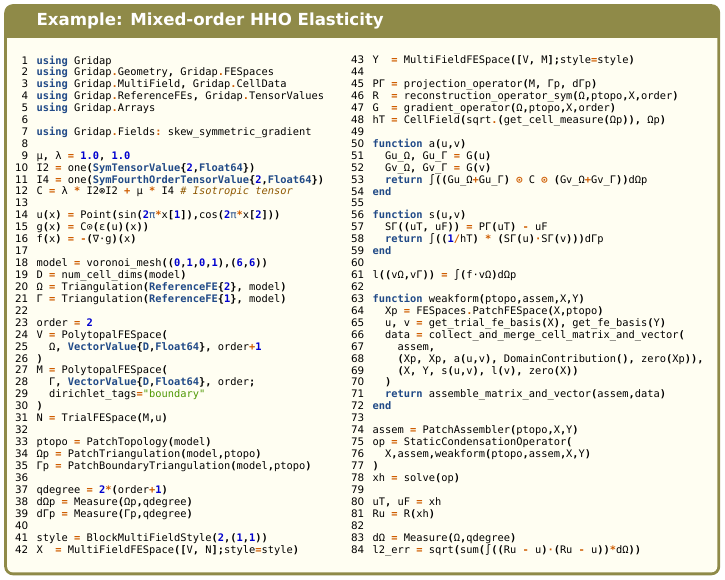}
  \caption{Mixed-order \ac{hho} discretisation of the linear elasticity problem (\secref{subsec:example-elasticity}). Requires {the code in} \figref{fig:voronoi} and {the code in} \lstref{lst:lop_elasticity}.}
  \label{lst:example_mhho_elasticity}
\end{listing}
\begin{listing}[ht!]
  \juliacode{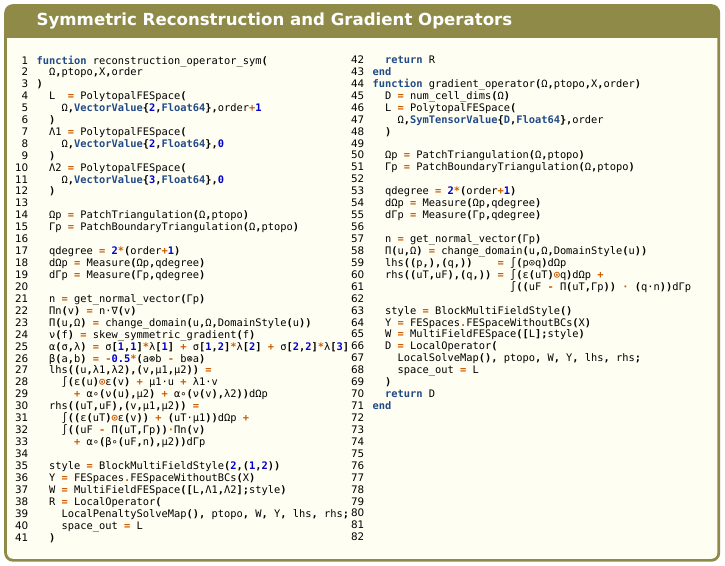}
  \caption{\ac{hho} local operators for the linear elasticity problem (\secref{subsec:example-elasticity}): symmetric reconstruction and gradient operators.}
  \label{lst:lop_elasticity}
\end{listing}


\subsubsection{\ac{hho} for the incompressible Stokes problem} \label{subsec:example-stokes}

Another typical example of a vector-valued problem is the incompressible Stokes problem \cite{AghiliBoyavalDiPietro2015}, which models the flow of an incompressible viscous fluid at low Reynolds numbers. We will closely follow the formulation introduced in \cite[Chapter 8]{DiPietro-Droniou-2020-HHO-Book}. \lstref{lst:example_mhho_stokes} shows the implementation of this example.

For the velocity field, we take $\vUh^{k+1,k}$ like in \eqref{eq:mhho-space-vector} (lines 16-23), while for the pressure field we take the broken polynomial space $\Pk{k}(\Th)$, split into its mean-value and zero-mean value subspaces. As directly imposing a global zero-mean value constraint on the pressure space is not scalable, we instead consider a cell-wise zero-mean pressure space and enforce global zero-mean value via a Lagrange multiplier. This leads to the following pressure spaces (lines 25-30):
\begin{equation} \label{eq:hho-space-pressure}
  P_h = \hat{P}_h \oplus \bar{P}_h \oplus \Lambda \ , \quad \hat{P}_h = \left\{ q_h \in \Pk{k}(\Th) : (q_h,1)_T = 0 \ , \forall T \in \Th \right\} \ , \quad \bar{P}_h = \Pk{0}(\Th) \ , \quad \Lambda = \Pk{0}(\dom).
\end{equation}
The subspace $\hat{P}_h$ can be statically condensed together with the cell velocity unknowns, leaving only the mean-value pressure unknowns $\bar{P}_h$ and the Lagrange multiplier $\Lambda$ in the global system. Thus, we use \texttt{BlockMultiFieldStyle(2,(2,3))} (lines 32-34) to structure the five-field space as a two-block system $(\bs{U}_h \times \hat{P}_h, \Uhb \times \bar{P}_h \times \Lambda)$, where the first block contains all the variables that will be statically condensed.

To reconstruct the velocity field, we will use a vector-valued version of the reconstruction operator introduced in \eqref{eq:hho-reconstruction}. Given $\ul{v}_T\in \vUh(T)$, the local velocity reconstruction operator $\vecrec : \vUh^{k+1,k}(T) \to \left[\Pk{k+1}(T)\right]^d$ is uniquely defined by
\begin{align} \label{eq:hho-velocity-reconstruction-stokes}
  (\nabla \vecrec \ul{v}_T, \nabla \bm{w})_T &= - (v_T, \Delta \bm{w})_T + \sum_{F \in \Fh(T)} (v_F - v_T, \nabla \bm{w} \cdot \bm{n}_{TF})_F \quad \forall \bm{w} \in \left[\Pk{k+1}(T)\right]^d, \\
  (\vecrec \ul{v}_T, 1)_T &= (v_T, 1)_T
\end{align}
To reconstruct the divergence, we will introduce a new local operator $\divrec : \vUh^{k+1,k}(T) \to \Pk{k}(T)$ defined by
\begin{align} \label{eq:hho-divergence-reconstruction-stokes}
  (\divrec \ul{v}_T, q)_T &= (\nabla \cdot v_T, q)_T + \sum_{F \in \Fh(T)} {\left( (v_F - v_T)\cdot \bm{n}_{TF}, q \right)_F} \quad \forall q \in \Pk{k}(T)
\end{align}
The implementation of these operators can be found in \lstref{lst:lop_stokes}. A subtle detail is that, when using these operators within \lstref{lst:example_mhho_stokes}, the input space of the operators is $\vUh$ (i.e. \texttt{YR} in lines 45-47) while the solution space of the problem is $\vUh \times P_h$. Thus, we need to recast the five-field variables into the velocity reconstruction operator's input space $\vUh$ (then recast the resulting bases back into the original space). To do so, we wrap our operators (lines 55-60) within a function \texttt{adapt\_ids} (lines 48-54). Alternatively, one could create a new local map, in the spirit of \lstref{lst:patch_assembly}, that is prepared to handle the five fields.

\begin{figure}[ht!]
  \centering
  \includegraphics[width=0.48\textwidth]{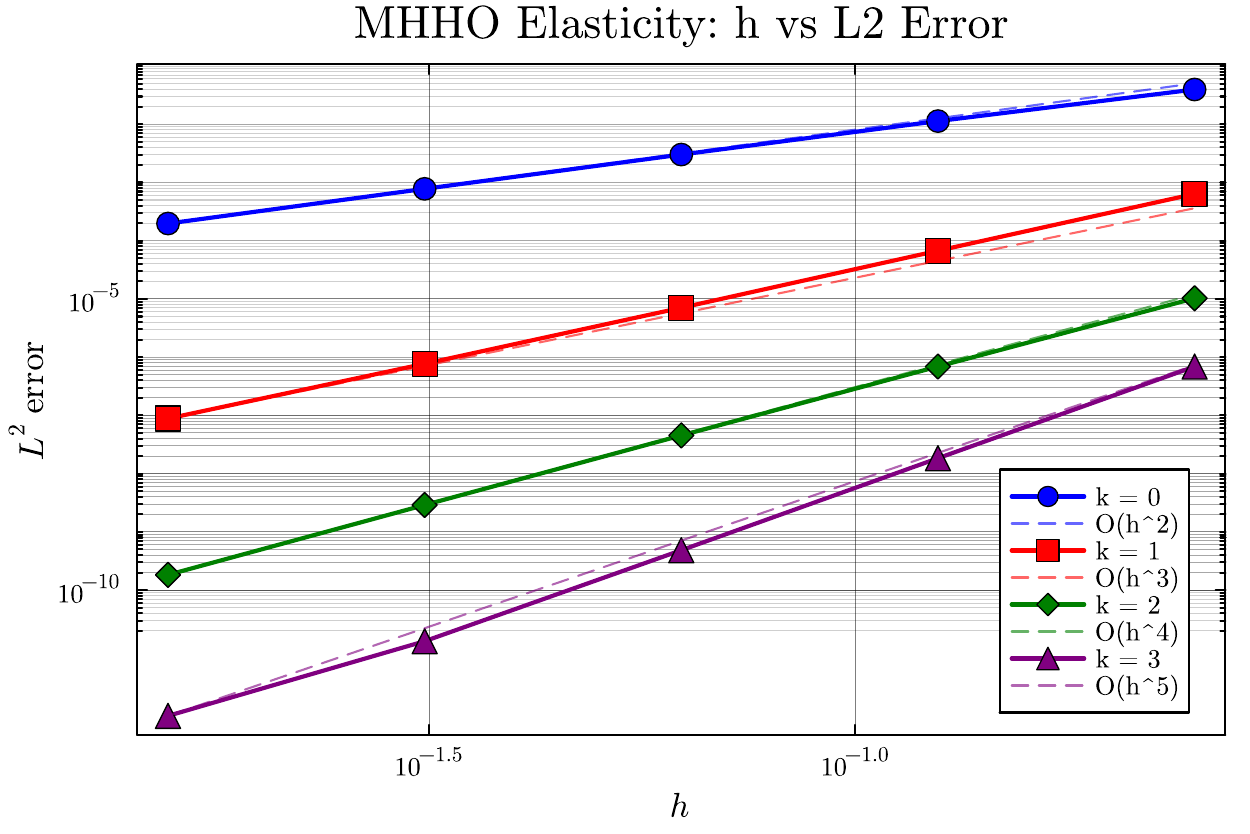}
  \hfill
  \includegraphics[width=0.48\textwidth]{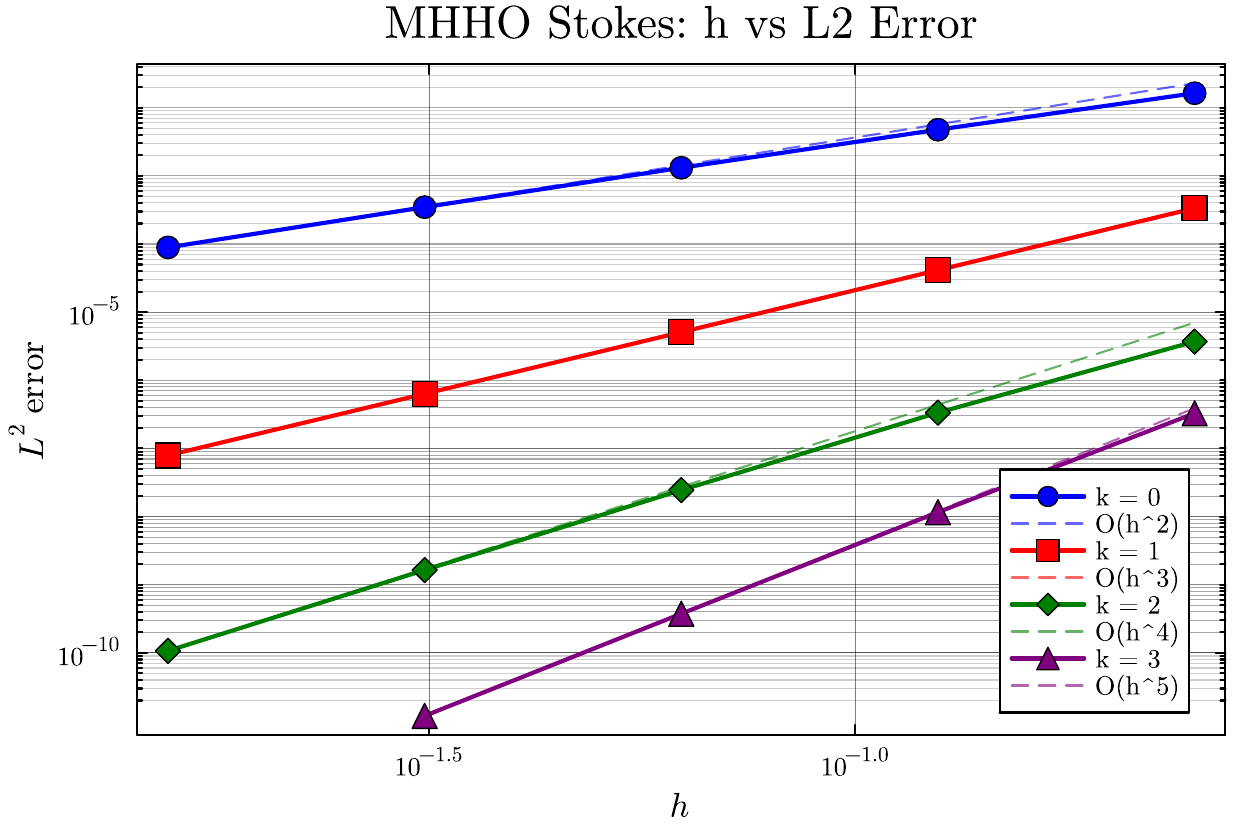}
  \caption{Mixed-order \ac{hho} method convergence: $L^2$-error versus mesh size $h$ for polynomial degrees $k=0,1,2,3$. Linear elasticity (left) and Stokes (right). Reference lines show the expected convergence rate $O(h^{k+2})$. Results generated on Voronoi meshes constructed as described in \figref{fig:voronoi}.}
  \label{fig:convergence-mhho-advanced}
\end{figure}

Going back to \lstref{lst:example_mhho_stokes}, the discretised problem reads (lines 65-99): find $(\uh, p_h) \in \vUh^{k+1,k} \times P_h$ such that $a_h((\uh, p_h); (\vh, q_h)) = l_h(\vh) \quad \forall (\vh, q_h) \in \vUh^{k+1,k} \times P_h$, where the local forms are defined as follows:
\begin{align*}
  &\hspace{1.5cm}a_T(\uT, p_T; \vT, q_T) = \int_T \left( \nabla \vecrec \ul{u}_T : \nabla \vecrec \ul{v}_T - p_T \divrec \ul{v}_T - q_T \divrec \ul{u}_T \right), \\
  &S_T(\uT,\vT) = \sum_{F \in \Fh(T)} h_F^{-1} \int_F \left( (\lproj{F}{0,k}u_T - u_F) \cdot (\lproj{F}{0,k}v_T - v_F) \right), \quad \text{and} \quad
  l_T(\vT) = \int_T \bm{f} \cdot v_T \, dx.
\end{align*}
Convergence results for this method can be found in \figref{fig:convergence-mhho-advanced}.
\begin{listing}[ht!]
  \juliacode{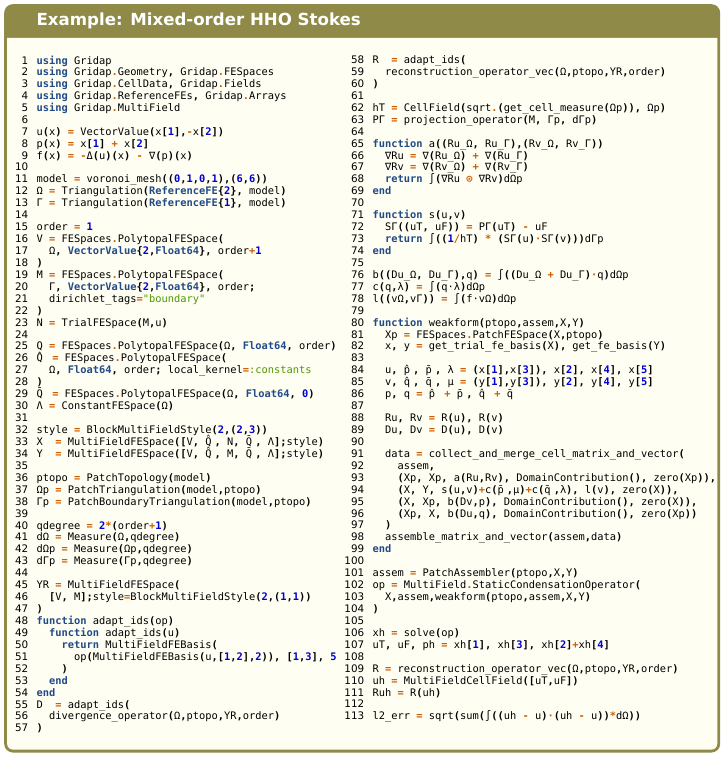}
  \caption{Mixed-order \ac{hho} discretisation of the incompressible Stokes problem (\secref{subsec:example-stokes}). Requires {the code in} \figref{fig:voronoi} and {the code in} \lstref{lst:lop_stokes}.}
  \label{lst:example_mhho_stokes}
\end{listing}
\begin{listing}[ht!]
  \juliacode{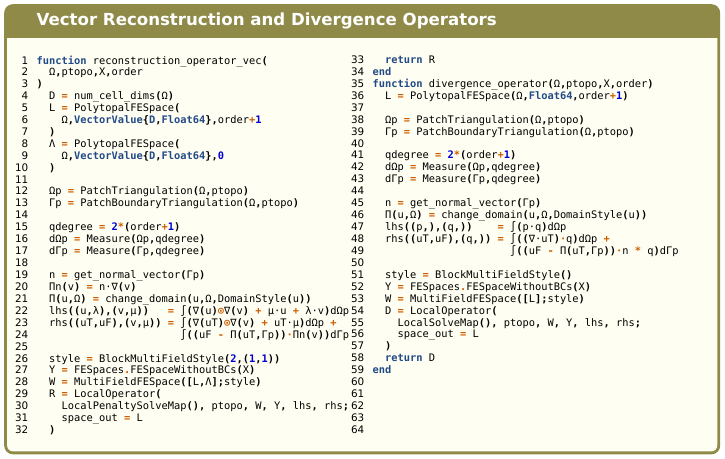}
  \caption{\ac{hho} local operators for the incompressible Stokes problem (\secref{subsec:example-stokes}): velocity reconstruction and divergence operators.}
  \label{lst:lop_stokes}
\end{listing}


\subsubsection{Optimal control}\label{sec:ex.oc} 
\begin{figure}[ht!]
  \centering
  \includegraphics[width=0.32\textwidth]{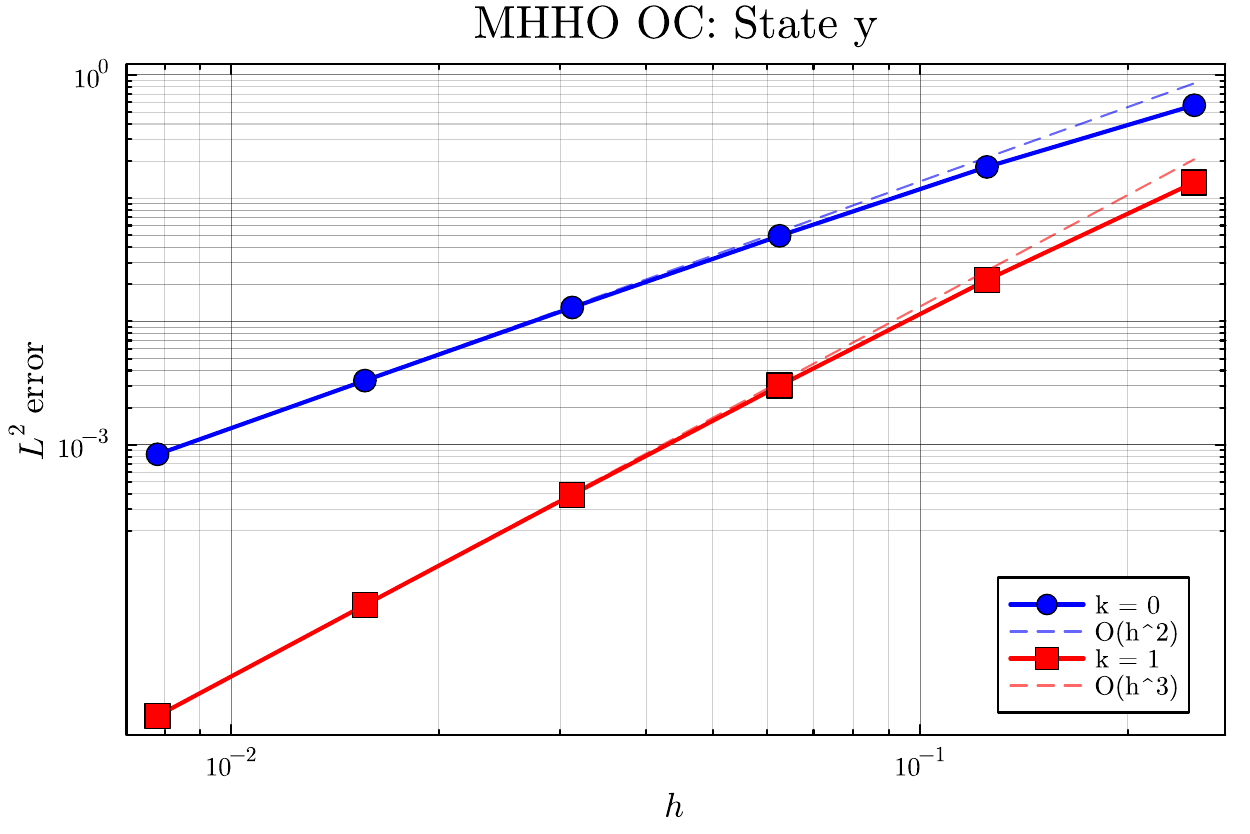}
  \hfill
  \includegraphics[width=0.32\textwidth]{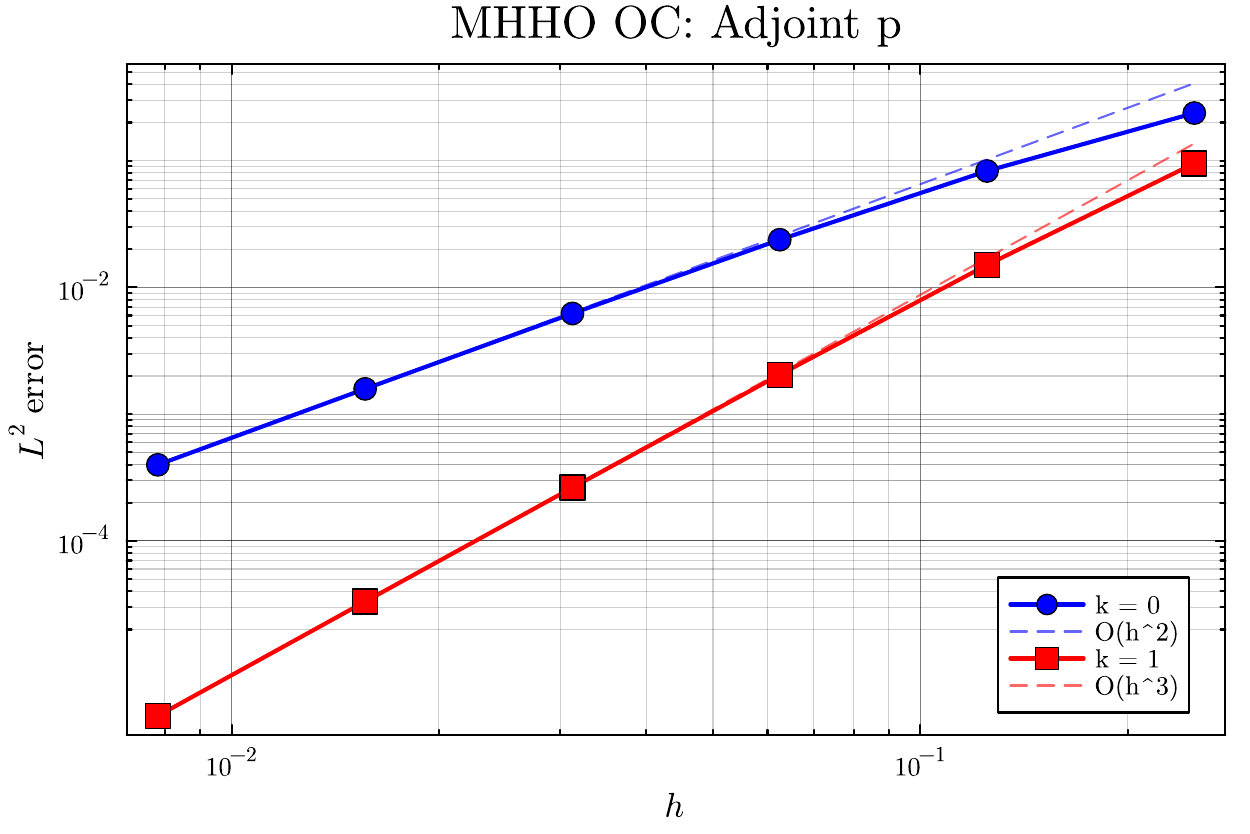}
  \hfill
  \includegraphics[width=0.32\textwidth]{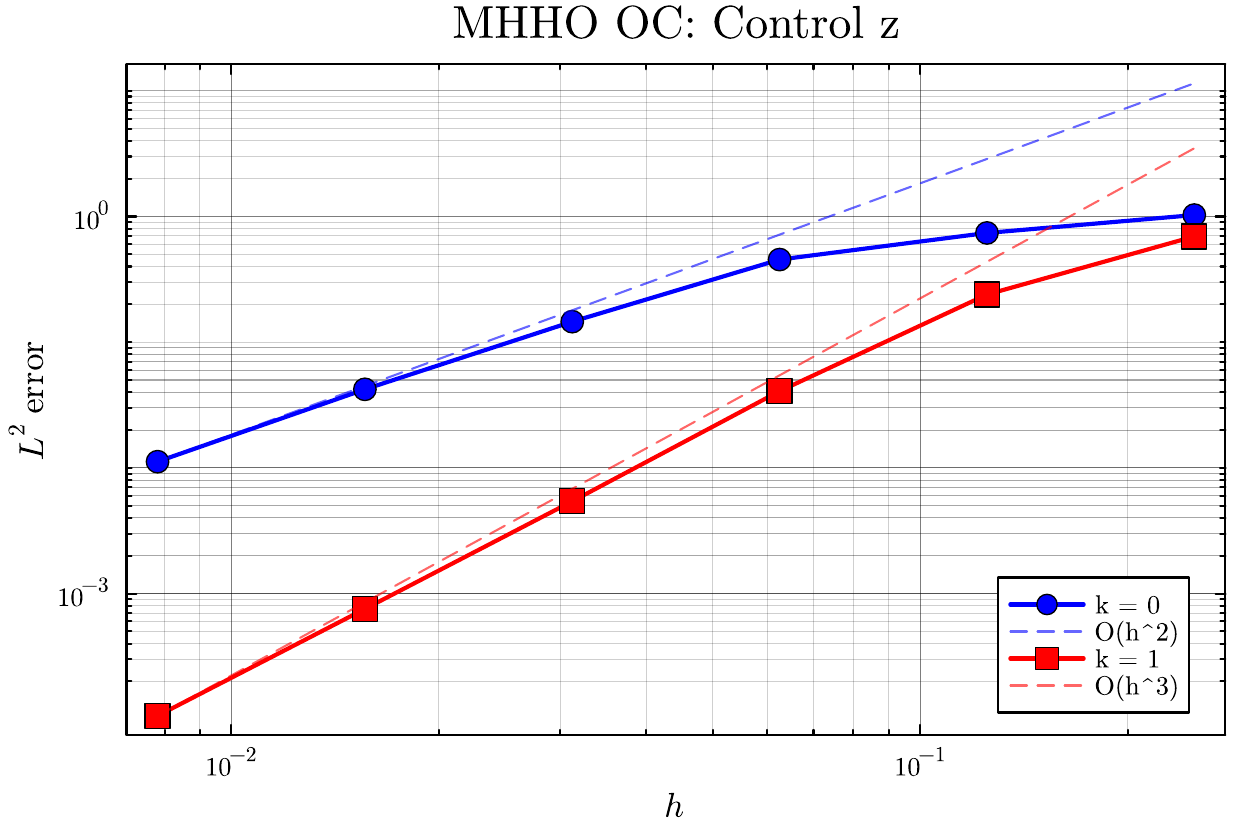}
  \caption{Mixed-order \ac{hho} method convergence for the optimal control problem: $L^2$-error versus mesh size $h$ for polynomial degrees $k=0,1$. State variable $u$ (left), adjoint variable $p$ (center) and control variable $z$ (right). Reference lines show the expected convergence rate $O(h^{k+2})$. Results generated using the driver from \lstref{lst:example_oc_poisson} on Voronoi meshes constructed as described in \figref{fig:voronoi}.}
  \label{fig:convergence-mhho-oc}
\end{figure}

Optimal control problems governed by \acp{pde} involve finding a control function that minimises a cost functional while satisfying the constraints imposed by the \ac{pde}. We consider a prototypical example of a distributed optimal control problem governed by the Poisson equation \cite{Troeltzsch2010}. Given a bounded domain $\dom \subset \mathbb{R}^2$, with boundary $\partial \dom$, we seek to minimise the cost functional
\begin{align*}
\min_{z \in Z_{ad}} J(u, z) = \frac{1}{2} \| u - u_d \|_{L^2(\dom)}^2 + \frac{\alpha}{2} \| z \|_{L^2(\dom)}^2,
\end{align*}
subject to the state equation
\begin{align*}
  -\Delta u = z + f \quad \text{in} \; \dom, \quad u = 0 \quad \text{on} \; \partial \dom, \quad  z \in Z_{ad} := \{ z \in L^2(\dom): z_a \leq z(x) \leq z_b \text{ a.e. in } \dom \},
\end{align*}
where $u$ is the state variable, $z$ is the control variable, $u_d$ is the desired state, $\alpha > 0$ is a regularisation parameter and $z_a < z_b$ are real numbers. It is well known that the above distributed optimal control problem is equivalent to solving the following \ac{kkt} optimality system: $(u,p,z) \in H_0^1(\dom) \times H_0^1(\dom) \times Z_{ad}$
\begin{align*}
	 a(u,v) = (z + f, v), \quad a(v,p) = (u - u_d, v), \quad (p + \alpha z, w - z) \geq 0,
\end{align*} 
for all $v,w \in H_0^1(\dom)$ and $w \in Z_{ad}$. Here, $p$ is the adjoint variable which solves the corresponding adjoint equation.
We approximate the \ac{kkt} system using a mixed order \ac{hho} method using the implicit characterisation of the control variable $z = \min\{ z_b, \max\{z_a, -(1 / \alpha) p\}\}$. The discrete \ac{kkt} system reads: find $(\uh, \ul{p}_h) \in \Uh^{k+1,k} \times \Uh^{k+1,k}$ such that
\begin{align*}
	 a_h(\uh, \vh) = l_h(v_{\Th}) + (\min\{ z_b, \max\{z_a, -(1/\alpha) p_{\Th}\}\}, v_{\Th}), \quad a_h(\ul{w}_h, \ul{p}_h) = (w_{\Th}, u_{\Th} - u_d),
\end{align*}
for all $\vh, \ul{w}_h \in \Uh^{k+1,k}$. The optimal control can be post-processed using the characterisation $z_{\Th} = \min\{ z_b, \max\{z_a, (-1/ \alpha) p_{\Th}\}\} \in Z_{ad}$. The following code snippet (see \lstref{lst:example_oc_poisson}) shows the implementation of this example.
\begin{listing}[]
  \juliacode{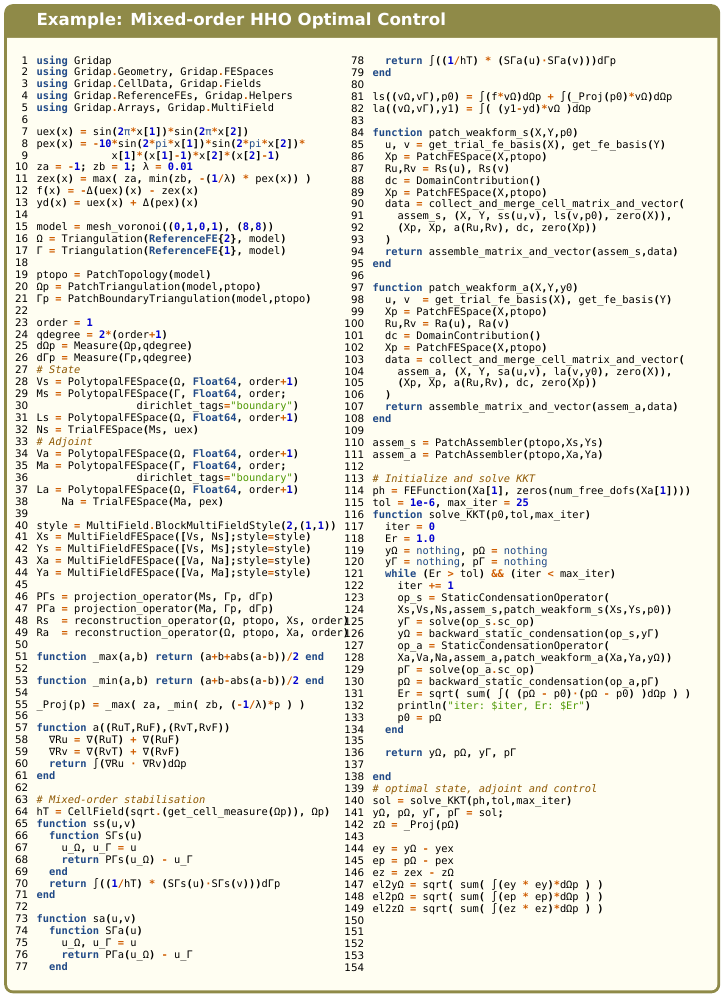}
  \caption{Mixed-order \ac{hho} discretisation of the optimal control problem (\secref{sec:ex.oc}). Requires {the code in} \figref{fig:voronoi} and {the code in} \lstref{lst:lop_poisson}.}
  \label{lst:example_oc_poisson}
\end{listing}
Convergence results for this method can be found in \figref{fig:convergence-mhho-oc}.

\section{Conclusion}

We have presented a comprehensive framework for implementing non-conforming hybrid polytopal \ac{fe} methods within the Gridap library. By introducing abstractions for polytopal mesh representation, broken polynomial spaces, patch-based local assembly, local operators and static condensation, we enable researchers to implement complex hybrid methods with minimal code while maintaining computational efficiency.

The framework's design leverages Julia's multiple dispatch and Gridap's lazy evaluation to achieve a high-level declarative programming model that closely resembles mathematical notation. Complete implementations of \ac{dg}, \ac{hdg} and \ac{hho} methods require around 100 lines of code, dramatically reducing the implementation burden compared to traditional approaches. Despite this brevity, the generated code achieves performance comparable to hand-optimised implementations through \ac{jit} compilation. 
Although not discussed in this work, the framework also supports distributed-memory parallelism through GridapDistributed \cite{GridapDistributed} and scalable solvers through GridapSolvers \cite{GridapSolvers}, enabling large-scale simulations on polytopal meshes \cite{badia2025analysis,badia2026gmg}. 

The numerical examples presented in \secref{sec:examples} demonstrate that the framework successfully handles methods of varying complexity, from simple \ac{dg} discretisations to sophisticated \ac{hho} formulations involving multiple local reconstruction operators. The convergence results confirm that the implementations achieve optimal rates on polytopal meshes, validating both the mathematical formulations and the computational framework.

Beyond the specific methods demonstrated here, the abstractions introduced in this work are sufficiently general to support a broader class of polytopal discretisations. The patch assembly framework can accommodate overlapping patches and multi-cell configurations, enabling applications to patch-based multigrid smoothers and other advanced techniques. 

Future work will focus on extending the framework to support more variants of hybrid methods, such as \ac{vem} or the discrete de Rham sequence on polytopal meshes \cite{discrete_derham}.

\section*{Acknowledgments}
This research was partially funded by the Australian Government through the Australian Research Council (project numbers DP210103092 and DP220103160). 

\printbibliography

\end{document}